\documentclass[a4paper,10pt,aps,amsfonts,amsmath,nofootinbib,byrevtex,prd,notitlepage]{revtex4-1}

\usepackage[UKenglish]{babel}
\usepackage{drcmath}
\usepackage{graphicx}
\usepackage{slashed}

\DeclareVector{\valpha}{\alpha}

\begin{document}

\title{Variational and Dyson--Schwinger Equations of Hamiltonian Quantum Chromodynamics}

\author{Davide Campagnari}
\author{Hugo Reinhardt}
\affiliation{Institut f\"ur Theoretische Physik, Universit\"at T\"ubingen,
Auf der Morgenstelle 14, 72076 T\"ubingen, Germany}
\date{\today}

\begin{abstract}
The variational Hamiltonian approach to Quantum Chromodynamics
in Coulomb gauge is investigated within the framework of the canonical recursive
Dyson--Schwinger equations. The dressing of the quark
propagator arising from the variationally determined non-perturbative kernels is expanded
and renormalized at one-loop order, yielding a chiral condensate compatible with the observations.
\end{abstract}

\maketitle


\section{Introduction}

Confinement and spontaneous chiral symmetry breaking are the basic features of Quantum
Chromodynamics (QCD) at ordinary density and temperature. Chiral symmetry breaking is
responsible for almost the entire mass of the visible matter in the universe. Both phenomena
originate in the low-energy sector of the theory, where perturbation theory cannot be
applied, and are strongly intertwined: lattice calculations show evidence that the
deconfinement transition and restoration of chiral symmetry coincide, at least for
fermions in the fundamental representation. Thanks to intensive studies both on the
lattice \cite{Aoki:2016frl,Faber:2017alm} and in the continuum theory
\cite{Alkofer:2000wg,Pawlowski:2005xe,Fischer:2006ub,Binosi:2009qm,Eichmann:2016yit,Feuchter:2004mk,*Reinhardt:2004mm,Epple:2006hv,Vastag:2015qjd}, we
have gained essential insights into the basic features of the QCD vacuum, although a
rigorous understanding of both phenomena is still missing. From these studies three
pictures have emerged: the dual Mei\ss ner effect \cite{Nambu:1974zg,tHooft:1981bkw},
the centre vortex scenario \cite{Mack:1978rq,Nielsen:1979xu,DelDebbio:1998luz,Langfeld:1997jx},
and the Gribov--Zwanziger picture \cite{Gribov:1977wm,Zwanziger:1998ez}. Both lattice
and continuum studies have also shown that these pictures are related \cite{Greensite:2004ur,Reinhardt:2008ek,Burgio:2015hsa}.

The Gribov--Zwanziger picture emerged in the variational Hamiltonian approach to QCD
in Coulomb gauge \cite{Feuchter:2004mk,Reinhardt:2004mm,Epple:2006hv}:
indeed a confining quark potential is found, together with an infrared diverging ghost
form factor and gluon energy. A simplified variational calculation
\cite{Finger:1981gm,Amer:1983qa,LeYaouanc:1983huv,Adler:1984ri}
based on a BCS-type wave functional for the quark sector of QCD, in which the coupling
of the quarks to the transverse spatial gluons is neglected, shows that the confining
quark potential also triggers chiral symmetry breaking, although the corresponding order
parameter, i.e.~the chiral condensate, turns out to be too small. This model has been
phenomenologically improved by using more general two-body interactions \cite{Bicudo:1989sh,Alkofer:1988tc}.

In Ref.~\cite{Pak:2011wu,*Pak:2013uba} the variational Hamiltonian approach to
Yang--Mills theory in Coulomb gauge \cite{Feuchter:2004mk,Reinhardt:2004mm,Epple:2006hv}
was extended to full QCD by including the quark-gluon coupling explicitly in the vacuum
wave functional. The ansatz for the vacuum wave functional was further extended in
Refs.~\cite{Vastag:2015qjd,Campagnari:2016wlt} by adding a further Dirac structure to
the quark-gluon coupling in the trial vacuum wave functional. With this additional Dirac
structure the resulting gap equation is free of UV divergences. In the present paper we
show that this additional Dirac structure is also crucial to ensure multiplicative
renormalizability of the quark propagator. The results of Refs.~\cite{Vastag:2015qjd,Campagnari:2016wlt}
will be retraced here in the framework of the canonical recursive Dyson--Schwinger
equations (CRDSEs) \cite{Campagnari:2010wc,Campagnari:2015zsa}, which, in principle,
allows to go beyond the approximations used in Refs.~\cite{Vastag:2015qjd,Campagnari:2016wlt}
in a systematic way. The use of Dyson--Schwinger equations requires to formulate the
quark sector in terms of Grassmann variables, which turns out to be advantageous over
the operator formulation of Fock space used in Refs.~\cite{Vastag:2015qjd,Pak:2011wu}.
Besides reproducing the results of Refs.~\cite{Vastag:2015qjd,Campagnari:2016wlt} in the
framework of the CRDSEs we also investigate analytically the IR behaviour of the CRDSE
for the quark propagator and determine under which conditions quark confinement is realized.
Approximating the full quark-gluon vertex of the CRDSE by its bare counterpart we solve the variational
equation and investigate the one-loop renormalization of the quark propagator. From the
renormalized quark propagator we calculate the quark condensate.

The structure of this paper is as follows: In Sec.~\ref{sec:qcd} we reformulate the
Hamiltonian approach to QCD within the CRDSEs \cite{Campagnari:2015zsa} with the
vacuum wave functional proposed in Ref.~\cite{Vastag:2015qjd}. In Sec.~\ref{secwf}
we present the quark vacuum wave functional while in Sec.~\ref{subsec:qpdse} we derive
the corresponding CRDSEs for the quark propagator and the quark-gluon vertex, by
means of which the expectation value of the QCD Hamiltonian is evaluated. In Sec.~\ref{sec:IR}
we discuss the infrared behaviour of the dressing functions of the quark propagator
required for confinement. As an illustration of our approach in Sec.~\ref{sec:AD} 
we keep from the interaction of the quarks only the non-Abelian Coulomb potential resulting
in a massive extension of the model considered in Refs.~\cite{Finger:1981gm,Adler:1984ri}. In Sec.~\ref{sec:bare}
we show how to recover the results of Refs.~\cite{Vastag:2015qjd,Campagnari:2016wlt}
in the present approach by a leading-order skeleton expansion. In Sec.~\ref{sec:app}
we perform a semi-perturbative expansion of the quark propagator by using the variational
kernels as non-perturbative input and investigate the renormalizability of the quark
propagator. In Sec.~\ref{sec:m3m4} we discuss the relation between the mass function defined
in the four-dimensional quark propagator to the mass function of the three-dimensional
(equal-time) propagator. Some details concerning the coherent-state description of fermionic
states are given in the Appendix.

\section{QCD in Coulomb Gauge}\label{sec:qcd}

In Coulomb gauge the QCD Hamiltonian reads \cite{Christ:1980ku}
\begin{equation}\label{hh1}
\begin{split}
H ={}& \frac12 \int\d^3x \, J_A^{-1} \Pi_i^a(\vx) J_A \, \Pi_i^a(\vx) + \frac12 \int\d^3x \, B_i^a(\vx) \, B_i^a(\vx) \\
     &+ \int\d^3x \, \psi^\dag(\vx) \bigl[ -\I*\valpha\cdot\grad - g \valpha\cdot\vec{A}(\vx) + \beta m \bigr] \psi(\vx) \\
     &+ \frac{g^2}{2} \int \d^3x \d^3y \, J_A^{-1} \rho^a(\vx) \,J_A \, F_A^{ab}(\vx,\vy) \, \rho^b(\vy) ,
\end{split}
\end{equation}
where $\Pi_i^a=-\I*\delta/\delta A_i^a$ is the canonical momentum,
$B_i^a$ is the chromomagnetic field, $\psi$ and $\psi^\dag$
are the fermion field operators, $\alpha_i$ and $\beta$ are the usual Dirac matrices, $m$
is the bare current quark mass, and $\vec{A}=\vec{A}^{\!a\,} t^a$ are the (transverse) gauge fields with
$t^a$ being the hermitian generators of the $\mathfrak{su}(N_c)$ algebra.
The last term in \Eqref{hh1} is the so-called Coulomb term: it describes the interaction of the colour charge
density
\begin{equation}\label{hh2}
\rho^a(\vx) = \psi^\dag(\vx) \, t^a \psi(\vx) + f^{abc} A_i^b(\vx) \, \frac{\delta}{\I* \delta A_i^c(\vx)}
\end{equation}
through the Coulomb kernel
\begin{equation}\label{coulkernel}
F_A^{ab}(\vx,\vy) = \int\d^3z \, G_A^{ac}(\vx,\vz) \bigl( - \nabla^2_z \bigr) G_A^{cb}(\vz,\vy) ,
\end{equation}
where
\[
G_A^{-1}(\vx,\vy) = \bigl( - \delta^{ab} \nabla^2_x - g f^{acb} A_i^c(\vx) \partial_i^x \bigr) \delta(\vx-\vy)
\]
is the Faddeev--Popov operator of Coulomb gauge with $f^{acb}$ being the structure constants
of the $\mathfrak{su}(N_c)$ algebra. Finally, $J_A=\Det G_A^{-1}$  is the Faddeev--Popov
determinant of Coulomb gauge.

\subsection{\label{secwf}The Vacuum Wave Functional}

In the coherent-state description of the fermionic Fock space introduced in Ref.~\cite{Campagnari:2015zsa}
(see Appendix) a physical state $\ket{\varPsi}$ is described by a functional
\[
\braket{\xi,\xi^\dag,A}{\varPsi} = \varPsi[\xi_+^\dag,\xi_-,A]
\]
of the gauge fields $A_i$ and of the spinor-valued Grassmann fields
\begin{equation}\label{fn1}
\xi_\pm(1) = \Lambda_\pm(1,2) \xi(2),
\end{equation}
where
\begin{equation}\label{qptproj3}
\Lambda_\pm (1, 2) = \int \dfr[3]{p} \e^{\I* \vp \cdot (\vx_1 - \vx_2)} \Lambda_\pm (\vp) ,
\qquad \Lambda_\pm(\vp) = \frac12 \pm \frac{\valpha\cdot\vp+\beta m}{2 \sqrt{\vp^2+m^2}}
\end{equation}
are the projectors onto positive/negative energy eigenstates of the free Dirac operator
\begin{equation}\label{dir}
h_0(\vp) = \valpha\cdot\vp + \beta m .
\end{equation}
In coordinate space we employ a notation where a
numerical index stands collectively for the spatial coordinate as well as for the colour and
Lorentz indices. A repeated numerical index like in \Eqref{fn1} implies integration over the spatial coordinate
and summation over the discrete indices.
Matrix elements of an operator $O$ between physical states $\varPhi$ and $\varPsi$ are
given by the functional integral
\begin{equation}\label{evg1}
\begin{aligned}[b]
\bra{\varPhi} O\bigl[A,\Pi,\psi,\psi^\dag\bigr] \ket{\varPsi}
= {}&{} \int \calD \xi \calD \xi^\dag \calD A \, J_A \e^{-\mu} \, \varPhi^*[\xi_+^\dag,\xi_-,A] \\
&\times O \biggl[A,-\I \frac{\delta}{\delta A}, \xi_-+\frac{\delta}{\delta\xi_+^\dag}, \xi_+^\dag+\frac{\delta}{\delta\xi_-} \biggr]
\varPsi[\xi_+^\dag,\xi_-,A] ,
\end{aligned}
\end{equation}
where
\[
\mu = \xi^\dag(1) \, S_0(1,2) \, \xi(2) =  \int \dfr[3]p \, \xi^\dag(\vp) \, S_0(\vp) \, \xi(\vp) ,
\]
is the integration measure of the fermionic coherent states, which involves the bare
quark propagator
\begin{equation}\label{evg4}
S_0(\vp) = \frac{h_0(\vp)}{2E_\vp} = \frac{\valpha\cdot\vp + \beta m}{2 E_\vp}, \qquad E_\vp = \sqrt{\vp^2+m^2} .
\end{equation}

For the vacuum wave functional of QCD we take the ansatz
\begin{equation}\label{evg2}
\varPsi[A,\xi^\dag_+,\xi_-] \propto \exp\biggl\{ -\frac12 S_A[A] - S_f [\xi_+^\dag,\xi_-,A] \biggr\} ,
\end{equation}
where $S_A$ and $S_f$ define, respectively, the wave functionals of pure Yang--Mills
theory and of the quarks interacting with the gluons. We choose the latter in the form
\begin{equation}\label{ans2}
\begin{aligned}[b]
S_f [\xi_+^\dag,\xi_-,A] &= \xi_+^\dag(1) \bigl[ K_0(1,2) + K(1,2;3) A(3) \bigr] \xi_-(2) \\[\jot]
&=\xi^\dag(1') \Lambda_+(1',1) \bigl[ K_0(1,2) + K(1,2;3) A(3) \bigr] \Lambda_-(2,2') \xi(2')
\end{aligned}
\end{equation}
where $K_0$ and $K$ are variational kernels, whose form will be specified in more detail
later.

Once the functional derivatives in \Eqref{evg1} are taken, expectation values of
operators boil down to quantum averages of functionals of the fields
\[
\vev{f[\xi^\dag,\xi,A]} = \int \calD\xi^\dag \, \calD\xi \, \calD A \, J_A \e^{-S} \, f[\xi^\dag,\xi,A]
\]
with an ``action''
\begin{equation}\label{hh5}
S = S_A + S_f^{} + S_f^* + \mu .
\end{equation}
This equivalence between expectation values in the Hamiltonian approach and quantum
averages in the functional integral formulation of a Euclidean field theory in $d=3$ dimensions is the basis for the Dyson--Schwinger
approach \cite{Campagnari:2010wc,Campagnari:2015zsa} employed in this work. With the help of
familiar Dyson--Schwinger techniques the various one-particle irreducible equal-time
Green functions of the Hamiltonian approach can be related to the kernels
occurring in the ``action'' \Eqref{hh5}, i.e.~in the vacuum wave functional \Eqref{evg2},
by means of an infinite tower of integral equations.
These are named canonical recursive Dyson--Schwinger equations (CRDSEs) in order to make
clear that, while they look like standard DSEs, their physical content is somewhat different.
(The bare $n$-point vertices are not given by the action of the theory but by variational kernels
of the vacuum wave functionals.)

Notice that the variational kernels $K_0$ and $K_i$ in \Eqref{ans2} enter the
action \Eqref{hh5} (and therefore the CRDSEs) only in the combinations
\begin{equation}\label{bqk}
\bar\gamma(1,2) = \Lambda_+(1,1') K_0(1',2') \Lambda_-(2',2) + \Lambda_-(1,1') K_0^\dag(1',2') \Lambda_+(2',2)
\end{equation}
and
\begin{equation}\label{bqgv}
\bar\Gamma_0(1,2;3) = \Lambda_+(1,1') K(1',2';3) \Lambda_-(2',2) + \Lambda_-(1,1') K^\dag(1',2';3) \Lambda_+(2',2) .
\end{equation}
In the following we will refer to $\bar\gamma$ as the biquark kernel, and to $\bar\Gamma_0$
as the bare quark-gluon vertex.\footnote{%
   The bare quark-gluon vertex $\bar\Gamma_0$ entering the vacuum wave functional \Eqref{evg2}
   should be distinguished from the quark-gluon coupling in the QCD Hamiltonian \Eqref{hh1}.
   }

The choice of the variational kernels in \Eqref{ans2} is subject to a restriction: the vacuum wave
functional must be invariant under global colour
rotations. These are generated by the total charge operator
\[
Q^a = \int \d^3x \, \rho^a(\vx),
\]
i.e.~the spatial integral
of the colour charge density \Eqref{hh2}. Invariance under global colour rotations
generated by $Q^a$ implies that the wave functional \Eqref{evg2} (or, equivalently, the
quantities $S_A$ and $S_f$ occurring in its exponent) must be annihilated by $Q^a$, which
leads to the condition
\[
Q^a S_f = \int\d^3x \d^3y \, \xi_+^{\dag}(\vx)
\biggl\{ \comm{t^a}{K_0(\vx,\vy)} + \int\d^3z \Bigl( \comm{t^a}{K_i^b(\vx,\vy;\vz)} - \I* f^{abc} K_i^c(\vx,\vy;\vz) \Bigr) A_i^b(\vz) \biggr\} \xi_-(\vy) \overset{!}{=} 0 .
\]
In order to satisfy this condition the variational kernels should obey the colour structure
$K_0\sim\id$ and $K_i^a\sim t^a$. Furthermore, since the $\Lambda_\pm$ are orthogonal projectors
[see \Eqref{qptproj3}] it is obvious from \Eqref{bqk} that the variational kernel $K_0$
must possess non-trivial Dirac structures. 

In principle, $K_0$ could have the general form
\begin{equation}\label{hh3}
K_0(\vp) = \beta s_1(\vp) + \valpha\cdot\vp \, s_2(\vp) + \beta \, \valpha\cdot\vp \, s_3(\vp)
\end{equation}
with complex scalar functions $s_1$, $s_2$, $s_3$. The resulting biquark kernel
\Eqref{bqk} becomes in momentum space
\[
\bar\gamma(\vp) = \frac{\beta \vp^2 - m \valpha\cdot\vp}{E_\vp^2} \Re\{ s_1(\vp) - m s_2(\vp) - E_\vp s_3(\vp) \}
- \frac{\I*\beta\valpha\cdot\vp}{E_\vp} \Im\{s_1(\vp) - m s_2(\vp) - E_\vp s_3(\vp) \} .
\]
As one observes, the complex kernels $s_1$, $s_2$, and $s_3$ enter the biquark kernel $\bar\gamma$
(and therefore all vacuum expectation values of observables as well as the CRDSEs) never individually but only in the combination
\[
s_1(\vp) - m s_2(\vp) - E_\vp s_3(\vp) .
\]
It is hence sufficient to consider only one of them; moreover, in the chiral limit $m=0$
the scalar kernel $s_2$ drops out. Therefore, the general ansatz \Eqref{hh3} is not necessary.
Instead the relevant physics can be captured by the much simpler choice
\begin{equation}\label{ans5}
K_0(\vp) = \beta \, s(\vp) ,
\end{equation}
which leads to the biquark kernel [\Eqref{bqk}]
\begin{equation}\label{ans10a}
\bar\gamma(\vp) = - \valpha\cdot\uvp \, \frac{m \abs{\vp}}{E_\vp^2} \Re s(\vp) + \beta \, \frac{\vp^2}{E_\vp^2} \Re s(\vp)
- \I*\beta\valpha\cdot\uvp \, \frac{\abs{\vp}}{E_\vp} \Im s(\vp).
\end{equation}

For the vector kernel we choose the ansatz \cite{Vastag:2015qjd}
\begin{equation} \label{ans6}
K_i^{mn,a}(\vp,\vq;\vk) = g \, t^a_{mn} \, \bigl[ \alpha_i \, V(\vp,\vq) + \beta \, \alpha_i \, W(\vp,\vq) \bigr] (2\pi)^3 \delta(\vp + \vq + \vk) ,
\end{equation}
where $V$, and $W$ are variational kernels: For simplicity we write only their dependence
on the quark--anti-quark momenta, as momentum conservation implicitly fixes
the gluon momentum.
Note that the vectorial character of the quark-gluon coupling is entirely given by the
Dirac matrix $\alpha_i$, i.e.~the variational kernels $V$ and $W$ are scalar functions
which may depend only on $\vp^2$, $\vq^2$, and $\vp\cdot\vq$, implying e.g.~$V(-\vp,-\vq) = V(\vp,\vq)$.
The bare quark-gluon vertex \Eqref{bqgv} becomes with \Eqref{ans6} in the chiral limit
\begin{equation}\label{qgv0}
\begin{split}
\bar{\Gamma}_{0,i}^{mn,a}(\vp,\vq;\vk) ={}&
t^a_{mn} \, \frac{g}{4} \biggl[ (1+\valpha\cdot\uvp) \bigl[ V(\vp,\vq) \alpha_i + W(\vp,\vq) \beta \alpha_i \bigr](1+\valpha\cdot\uvq) \\
&+ (1-\valpha\cdot\uvp) \bigl[ V^*(\vq,\vp) \alpha_i - W^*(\vq,\vp) \beta \alpha_i \bigr](1-\valpha\cdot\uvq) \biggr]  (2\pi)^3 \delta(\vp + \vq + \vk).
\end{split}
\end{equation}

When both vector kernels are omitted, $V(\vp,\vq) = 0 = W(\vp,\vq)$, the wave functional
\Eqref{evg2} reduces to the BCS-type wave functional used in Refs.~\cite{Adler:1984ri,Finger:1981gm,Alkofer:1988tc},
while keeping only $V$ corresponds to the choice of Refs.~\cite{Pak:2011wu,*Pak:2013uba}.
The above ansatz for the fermionic wave functional defined by Eqs.~\eqref{evg2},
\eqref{ans2}, \eqref{ans5}, and \eqref{ans6} was also chosen in
Refs.~\cite{Vastag:2015qjd,Campagnari:2016wlt},
where the QCD variational principle was formulated in the ordinary operator language
of second quantization, avoiding the introduction of Grassmann fields. As shown in Ref.~\cite{Campagnari:2016wlt}
this ansatz has the advantage that all UV divergences cancel in the quark gap equation.

\subsection{Quark Propagator and Quark-Gluon Vertex CRDSEs}\label{subsec:qpdse}

\begin{figure}[tb]
\centering
\includegraphics[width=.5\linewidth]{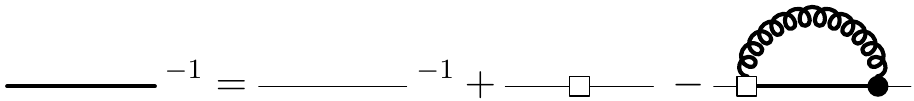}\\[2ex]
\includegraphics[width=.6\linewidth]{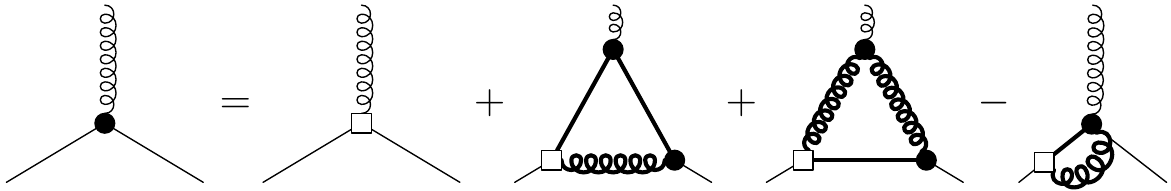}
\caption{Diagrammatic representation of the CRDSEs for the quark propagator [top, \Eqref{qdse2}]
and for the quark-gluon vertex [bottom]. Full lines and filled dots represent,
respectively, dressed propagators and vertices. The line with an empty square stands for the biquark
kernel $\bar\gamma$ [\Eqref{ans10a}]; the vertex with a square box represents the bare quark-gluon vertex [\Eqref{qgv0}].}
\label{fig-dse-quark}
\end{figure}

As shown in Refs.~\cite{Campagnari:2010wc,Campagnari:2015zsa} the formal equivalence between
expectation values in the Hamiltonian approach and quantum averages of a Euclidean field
theory can be used to write down DSE-like equations, referred to as CRDSEs to express the $n$-point functions
by the variational kernels of the vacuum wave functional. The CRDSE for the
fermion propagator
\[
Q(1,2) = \vev{\xi(1)\xi^\dag(2)}
\]
reads
\begin{equation}\label{qdse2}
Q^{-1}(1,2) = Q_0^{-1}(1,2) + \bar\gamma(1,2) - \bar\Gamma_0(1,3;4) Q(3,3') D(4,4') \bar\Gamma(3',2;4') ,
\end{equation}
where
\[
Q_0(1,2) = \Lambda_+(1,2) - \Lambda_-(1,2)
\]
is the bare fermion propagator,
\begin{equation}\label{gluonprop}
D(1,2) = \vev{A(1) A(2)}
\end{equation}
is the gluon propagator, and $\bar\Gamma$ is the full quark-gluon vertex defined by
\begin{equation}\label{qgv}
\vev{\xi(1) \xi^\dag(2) A(3)} = - Q(1,1') \, \bar\Gamma(1',2';3') \, Q(2',2) \, D(3',3) .
\end{equation}
The latter also obeys a CRDSE, which is represented diagrammatically together with \Eqref{qdse2}
in Fig.~\ref{fig-dse-quark}. The explicit form is not relevant for the present work but
the first term on the right-hand side is given indeed by $\bar\Gamma_0$ [\Eqref{qgv0}],
thus justifying its interpretation as bare quark-gluon vertex.

Equation \eqref{qdse2} may be conveniently written in momentum space: with the explicit
form \Eqref{ans10a} of the biquark kernel we obtain
\begin{equation}\label{qdse3}
\begin{split}
\bigl[Q^{mn}(\vp)\bigr]^{-1} ={}& \delta^{mn} \, \frac{\valpha\cdot\vp + \beta m}{E_\vp}
+ \delta^{mn} \biggl( - \valpha\cdot\vp \, \frac{m}{E_\vp^2} \Re s(\vp) + \beta \, \frac{\vp^2}{E_\vp^2} \Re s(\vp)
- \I*\beta\valpha\cdot\vp \, \frac{1}{E_\vp} \Im s(\vp)  \biggr) \\
&- \int \dfr[3]{q} \: \bar\Gamma_{0,i}^{mk,a}(\vp,-\vq;\vq-\vp) \, Q(\vq)\, D_{ij}(\vp-\vq) \, \bar\Gamma_{j}^{kn,a}(\vq,-\vp;\vp-\vq) .
\end{split}
\end{equation}
For the inverse quark propagator, which we assume to be colour diagonal, we must consider
in principle the following Dirac structure
\begin{equation}\label{qdse5}
Q^{-1}(\vp) = A(\vp) \, \valpha\cdot\uvp + \beta \,  B(\vp) - \I* \beta \valpha\cdot\uvp \, C(\vp) + D(\vp) ,
\end{equation}
which can be inverted to give
\begin{equation}\label{qdse5a}
Q(\vp) = \frac{ A(\vp) \, \valpha\cdot\uvp + \beta \,  B(\vp) - \I* \beta \valpha\cdot\uvp \, C(\vp) - D(\vp)}{A^2(\vp)+B^2(\vp)+C^2(\vp)-D^2(\vp)} .
\end{equation}
From the CRDSE~\eqref{qdse3} we obtain the following system of coupled equations for the
dressing functions \Eqref{qdse5} of the quark propagator
\begin{equation}
\begin{split}
\label{1010-17a}
A (\vp) &= \frac{\abs{\vp}}{E_\vp}\biggl( 1 - \frac{m}{E_\vp} \Re s\vp)\biggr)
- \frac{1}{4 \Nc} \int\dfr[3]{q} \: \tr[ \valpha\cdot\uvp \bar\Gamma_{0,i}^{mn,a}(\vp,-\vq;\vq-\vp) \, Q(\vq)\, D_{ij}(\vp-\vq) \, \bar\Gamma_{j}^{nm,a}(\vq,-\vp;\vp-\vq) ] , \\
B (\vp) &=  \frac{m}{E_\vp} + \frac{\vp^2}{E_\vp^2} \Re s(\vp)
- \frac{1}{4\Nc} \int\dfr[3]{q} \: \tr[\beta \bar\Gamma_{0,i}^{mn,a}(\vp,-\vq;\vq-\vp) \, Q(\vq)\, D_{ij}(\vp-\vq) \, \bar\Gamma_{j}^{nm,a}(\vq,-\vp;\vp-\vq) ], \\
C (\vp) &= \frac{\abs{\vp}}{E_\vp} \Im s(\vp)
- \frac{1}{4\Nc} \int\dfr[3]{q} \: \tr[ -\I\beta \, \valpha\cdot\uvp\bar\Gamma_{0,i}^{mn,a}(\vp,-\vq;\vq-\vp) \, Q(\vq)\, D_{ij}(\vp-\vq) \, \bar\Gamma_{j}^{nm,a}(\vq,-\vp;\vp-\vq) ], \\
D (\vp) &= - \frac{1}{4 \Nc} \int\dfr[3]{q} \: \tr[ \bar\Gamma_{0,i}^{mn,a}(\vp,-\vq;\vq-\vp) \, Q(\vq)\, D_{ij}(\vp-\vq) \, \bar\Gamma_{j}^{nm,a}(\vq,-\vp;\vp-\vq) ] ,
\end{split}
\end{equation}
where
\[
D_{ij}(\vp) \equiv \frac{t_{ij}(\vp)}{2 \Omega(\vp)} , \qquad t_{ij}(\vp) = \delta_{ij} - \frac{p_i p_j}{\vp^2}
\]
is the gluon propagator \Eqref{gluonprop}, conveniently parametrized in terms of the quasi-gluon energy $\Omega(\vp)$.

At this point it should be mentioned that the fermion propagator $Q$ is not the physical
quark propagator, which in the Hamiltonian approach is defined by
\[
S(1,2) = \frac12 \vev[\big]{\comm{\psi(1)}{\psi^\dag(2)}} .
\]
The commutator arises from the equal-time limit of the time-ordered operator product in
the full time-dependent theory. The quark propagator $S$ and the propagator $Q$ are related by
\begin{equation}\label{qprop}
S(\vp) = Q(\vp) - S_0(\vp) ,
\end{equation}
with $S_0(\vp)$ being the free quark propagator, \Eqref{evg4}. As long as no confusion is
possible we will keep referring indiscriminately to both $S(\vp)$ and $Q(\vp)$ as quark
propagator.

\subsection{The QCD Vacuum Energy Density}

The vacuum expectation value of the QCD Hamiltonian has been evaluated in Ref.~\cite{Campagnari:2015zsa},
to which we refer the reader for the details of the calculation; here we will merely
quote the relevant contributions to the energy density $e\equiv \vev{H}/(V \cdot \Nc)$
in momentum space.
The Dirac Hamiltonian [second line in \Eqref{hh1}] yields
\begin{equation}\label{eD2}
\begin{split}
e^{}_\mathrm{D} ={}& - \int \dfr[3]{q} \tr\bigl[ (\valpha\cdot\vq + \beta m) Q(\vq) \bigr] \\
&- g C_F \int \dfr[3]{q} \dfr[3]{\ell} \, D_{ij}(\vq+\vl) \,
\tr\bigl[ \alpha_i \, Q(\vq) \bar\Gamma_j(\vq,\vl) Q(-\vl) \bigr] \equiv e^{(0)}_\mathrm{D}+ e^{(1)}_\mathrm{D},
\end{split}
\end{equation}
where $C_F=(\Nc^2-1)/(2\Nc)$ is the quadratic Casimir invariant of the fundamental
representation of the $\mathfrak{su}(\Nc)$ algebra. The fermionic contribution to
the kinetic energy of the gluons [first term on the right-hand side of \Eqref{hh1}] is
given by
\begin{equation}\label{eE7}
e_E^q = -\frac{C_F}{8} \int\dfr[3]{q} \dfr[3]{\ell} \, t_{ij}(\vq+\vl)
\begin{aligned}[t]
\tr \bigl\{ & \bar\Gamma_{0,i}(\vq,-\vl) Q(\vl) \bar\Gamma_j(\vl,-\vq) Q(\vq) \\
&- Q_0(\vq) \bar\Gamma_{0,i}(\vq,-\vl) Q(\vl) Q_0(\vl)\bar\Gamma_{0,j}(\vl,-\vq) Q(\vq)\bigr\} .
\end{aligned}
\end{equation}
For simplicity, in Eqs.~\eqref{eD2} and \eqref{eE7} we have omitted the dependence of
the vertex functions on the gluon momentum, which follows from the fermionic momenta
kept in the above equations by momentum conservation.
Furthermore, we have assumed that the propagators are colour diagonal and that the colour
structure of the full quark-gluon vertex is given by the generator $t^a$ as for the bare vertex.
Finally, the Coulomb interaction of the fermionic charges reads
\begin{equation}\label{eCqq61}
e_\mathrm{C}^{qq} \simeq - g^2 \frac{C_F}{2} \int \dfr[3]{q} \dfr[3]{\ell} \: F(\vq-\vl) 
\tr\bigl\{ \bigl[Q(\vl)-\tfrac12 Q_0(\vl)\bigr] \bigl[Q(\vq)-\tfrac12 Q_0(\vq)\bigr] - \tfrac14 \bigr\} .
\end{equation}
Here, $F(\vp)$ is the expectation value of the Coulomb kernel \Eqref{coulkernel}, which
in the following calculations will be approximated by the simple form \cite{Epple:2006hv}
\begin{equation}\label{coulpot}
g^2 F(\vp) = \frac{8\pi\sigma^{}_\mathrm{C}}{\vp^4} + \frac{g^2}{\vp^2} ,
\end{equation}
with $\sigma^{}_\mathrm{C}$ being the Coulomb string tension.

Since the expectation value $e_\mathrm{D}^{(0)}$ of the single-particle Hamiltonian [first term in \Eqref{eD2}]
and the Coulomb interaction \Eqref{eCqq61} do not depend on the full quark-gluon vertex,
the Dirac traces can be worked out explicitly, yielding respectively
\begin{equation}\label{eDn1}
e^{(0)}_\mathrm{D} = - 4 \int\dfr[3]{q} \, \frac{\abs{\vq} \, A(\vq) + m  B(\vq)}{\Delta(\vq)}
\end{equation}
and
\begin{equation}\label{eCqq6}
\begin{aligned}[b]
e_\mathrm{C}^{qq} = - g^2 \frac{C_F}{2} \int \dfr[3]{q} \dfr[3]{\ell} \,
\frac{F(\vq-\vl)}{\Delta(\vq) \, \Delta(\vl)} 
\Bigl\{&
4 \bigl[ B(\vq) \, B(\vl) + D(\vq) \, D(\vl) \bigr] -\Delta(\vq) \Delta(\vl)\\
& + \uvq\cdot\uvl \, \Bigl[ 4 C(\vq)\,C(\vl) + \bigl(2A(\vq)-\Delta(\vq)\bigr) \bigl(2A(\vl)-\Delta(\vl)\bigr) \Bigr] \Bigr\} ,
\end{aligned}
\end{equation}
where we have introduced the abbreviation
\begin{equation}\label{qdse8}
\Delta(\vq) = A^2(\vq)+B^2(\vq)+C^2(\vq)-D^2(\vq)
\end{equation}
for the denominator of the quark propagator \Eqref{qdse5a}.


\section{\label{sec:IR}Infrared Behaviour of the Dressing Functions}

Before we proceed to derive the equations of motion of our variational approach by
minimizing the energy density with respect to the variational kernels, we discuss here
which conditions the dressing functions $A(\vp)$,~\ldots,~$D(\vp)$ of the quark propagator
\Eqref{qdse5} must satisfy in order to guarantee confinement and chiral symmetry breaking.
For given variational kernels of the wave functional these dressing functions are determined
by the quark propagator CRDSE \eqref{1010-17a}, while the variational kernels themselves
are determined by minimizing the energy density.

For simplicity we assume that the vector kernels $V(\vp,\vq)$ and $W(\vp,\vq)$ are real
and symmetric, and that the scalar kernel $s(\vp)$ is real (we can always restrict our
variational ansatz to these class of kernels): then, consistent solutions with
$D(\vp) = 0$ and $C(\vp) = 0$ exist, see Eqs.~\eqref{qdse6} below. We will furthermore
restrict our considerations to chiral quarks, $m=0$.

As we have shown in Sec.~\ref{subsec:qpdse}, the physical quark propagator $S$
is related to the propagator $Q$ of the Grassmann fields by \Eqref{qprop} and can be
expressed through the dressing functions $A$ and $B$ as
\begin{equation}\label{prop3}
S(\vp) = Q(\vp) - S_0(\vp)
= \frac{[A_p(2-A_p)-B_p^2] \valpha\cdot\uvp + 2 B_p \beta}{2(A_p^2+B_p^2)} .
\end{equation}
In order to prevent the notation from becoming excessively cluttered we have expressed the momentum
dependence of the dressing functions through a subscript. 

Inspired by the form of the bare quark propagator [\Eqref{evg4}] we define the running
mass $M_p$ and the and the quark dressing function $Z_p$ by
\begin{equation}\label{prop4}
S(\vp) = Z_p \, \frac{\valpha\cdot\vp + \beta M_p}{2 \calE_p} , \qquad \calE_p=\sqrt{p^2+M_p^2} .
\end{equation}
From Eqs.~\eqref{prop3} and \eqref{prop4} we obtain
\begin{equation}\label{massf1}
M_p = \frac{2 p B_p}{A_p(2-A_p)-B_p^2}, \qquad
Z_p = \frac{\sqrt{[A_p(2-A_p)-B_p^2]^2+4B_p^2}\vphantom{B_p^2}}{A_p^2+B_p^2} ,
\end{equation}
where $p=\abs{\vp}$.
These equations can be inverted to express the dressing functions $A_p$ and $B_p$ in terms
of $M_p$ and $Z_p$ as
\begin{equation}\label{dress}
A_p = \frac{2 (\calE_p + p Z_p)}{\calE_p(1+Z_p^2)+2pZ_p} , \qquad
B_p = \frac{2 M_p Z_p }{\calE_p(1+Z_p^2)+2pZ_p} .
\end{equation}
Note that the approximation $A_p=1$ is equivalent to $Z_p=1$.
We will now exploit these relations to investigate the IR behaviour of the dressing
functions $A_p$ and $B_p$.

An IR finite mass function $M (p = 0)\equiv M_0 \neq 0$ is an indicator of chiral symmetry breaking. Therefore we investigate now
which conditions the functions $A_p$ and $B_p$ must fulfil at vanishing momentum so that
$M_0\neq0$. From \Eqref{massf1} follows immediately that an IR diverging $B_p$ and an IR finite
$A_p$ would give rise to a vanishing (negative!) mass function. The dressing function $B_p$
must therefore have an finite IR limit $B_0$. Furthermore, from the first equation in (\ref{massf1}) 
follows that the dressing function $A_p$ must also have
an finite IR limit $A_0$ satisfying the condition
\begin{equation}\label{massf2}
A_0(2-A_0) = B_0^2 \, .
\end{equation}
Hence for real $B_0$ and $A_0$ we find that $A_0\in[0,2]$. From the second expression
in \Eqref{massf1} we find in the limit of vanishing momentum assuming that \Eqref{massf2} holds
\begin{equation}\label{irqen1bis}
Z_0 = \sqrt{\frac{2-A_0}{A_0}} .
\end{equation}
Like \Eqref{massf2}, the right-hand side of \Eqref{irqen1bis} is well defined only for $A_0\in[0,2]$.
An infrared suppressed propagator $Z_0<1$ requires $A_0>1$, and an IR vanishing quark
propagator requires $A_0=2$, which in view of \Eqref{massf2} implies $B_0=0$. For the mass function to be
still non-vanishing in the IR, the dressing function $A$ should have zero slope at vanishing
momentum, as it can be seen by Taylor expanding \Eqref{massf1}.

From this IR analysis there emerges a possible Gribov--Zwanziger-like scenario which
includes both confinement and chiral symmetry breaking: an IR vanishing dressing function
$B_p$ and a dressing function $A_p$ satisfying $A(0)=2$ and $A'(0)=0$ yield an IR finite
running mass (i.e.~spontaneous breaking of chiral symmetry) and an IR vanishing (i.e.~confined)
quark propagator.
The same conclusions follow of course from \Eqref{dress} taken at zero momentum
\[
A_0 = \frac{2}{1+Z_0^2} , \qquad
B_p = \frac{2 Z_0}{1+Z_0^2} .
\]
For an infrared vanishing quark propagator, $Z_0=0$, we find immediately $A_0=2$ and $B_0=0$.

The above results are based in the analysis of the unrenormalized CRDSEs and may hence
change after renormalization. However, the renormalization affects mostly the UV behaviour.


\section{\label{sec:AD}Massive Adler--Davis Model}

To make contact with previous work and for the sake of illustration, in the present
section let us neglect the
quark-gluon coupling in the QCD Hamiltonian and consider the quark sector only. The
remaining contributions to the energy density are therefore Eqs.~\eqref{eDn1} and 
\eqref{eCqq6}. If we neglect the coupling of the
quarks to the transverse (spatial) gluons in the vacuum wave functional \Eqref{evg2},
\eqref{ans2}, i.e.~$V=0=W$, the bare quark-gluon vertex
\Eqref{qgv0} vanishes, $\bar\Gamma_0=0$. Furthermore, if the scalar kernel $s_p$ is real
both dressing functions $C_p$ and $D_p$ vanish identically. Then the energy density reduces to
\begin{equation}\label{ad3}
\begin{aligned}[b]
e^{}_\mathrm{AD} ={}& - 4 \int\dfr[3]{q} \, \frac{\abs{\vq} \, A_q + m  B_q}{\Delta_q} \\
&- g^2 \frac{C_F}{2} \int \dfr[3]{q} \dfr[3]{\ell} \,
\frac{F(\vq-\vl)}{\Delta_q \, \Delta_\ell} 
\Bigl\{ 4 B_q \, B_\ell + \uvq\cdot\uvl \bigl[A_q(2-A_q)-B_q^2 \bigr] \bigl[A_\ell(2-A_\ell)-B_\ell^2 \bigr]\Bigr\} ,
\end{aligned}
\end{equation}
while the dressing functions \Eqref{1010-17a} of the quark propagator become
\begin{equation}\label{adn1}
A_p = \frac{\abs{\vp}}{E_p}\biggl( 1 - \frac{m}{E_p} s_p\biggr)  , \qquad
B_p =  \frac{m}{E_p} + \frac{\vp^2}{E_p^2} s_p .
\end{equation}
Inserting these expressions into \Eqref{ad3} yields
\[
e^{}_\mathrm{AD} = - 4 \int \dfr[3]{q} \, \frac{E_q}{1+w^2_q} 
+ g^2 C_F  \int \dfr[3]{q} \dfr[3]\ell \,
\frac{F(\vq-\vl)}{E_q \, E_\ell} \frac{(m+q w_q) (m w_\ell - \ell) (w_\ell - \uvq\cdot\uvl \, w_q)}{(1+w^2_q) (1+w^2_\ell)} .
\]
where we have introduced the abbreviation
\begin{equation}\label{ad5}
w_p = \frac{\abs{\vp} s_p}{E_p} .
\end{equation}
Variation of $e^{}_\mathrm{AD}$ with respect to $s_p$ (or, equivalently, with respect to $w_p$)
yields the gap equation
\begin{equation}\label{ad8}
E_p \, w_p = \frac{g^2 C_F}{2} \int \dfr[3]{q} \,F(\vp-\vq) \, \frac{p}{E_p} \frac{q}{E_q}
 \frac{\calA(\vp;\vq) - \uvp\cdot\uvq \, \calA(\vq;\vp)}{1+w^2_q} ,
\end{equation}
with
\begin{equation}\label{ad9}
\calA(\vp;\vq) = \Bigl[ w_q + \frac{m}{2 q} ( 1-w^2_q) \Bigr] \Bigl[1-w^2_p - 2 \frac{m}{p} w_p \Bigr] .
\end{equation}
Putting $m=0$ in Eqs.~\eqref{ad8} and \eqref{ad9} and approximating the Coulomb potential
$F(\vp)$ [\Eqref{coulpot}] by its infrared part $8\pi\sigma^{}_\mathrm{C}/p^4$ yields
precisely the gap equation obtained by Adler and Davis \cite{Adler:1984ri}.
Equations~\eqref{ad8} and \eqref{ad9} give the extension of their model to finite
current quark masses. The integral on the right-hand side of \Eqref{ad8} appears
also in Ref.~\cite{Alkofer:1988tc}, where a slightly extended
phenomenological model for the quark--quark interaction was considered.

From the dressing functions \Eqref{adn1} we can calculate the quark propagator $Q$
[\Eqref{qdse5a}]
\[
Q(\vp) = \frac{\valpha\cdot\uvp ( p- m \, w_p ) + \beta ( m + p \, w_p )}{E_\vp (1+w^2_p)} ,
\]
and after elementary but somewhat lengthy algebra
the true quark propagator $S$ [see \Eqref{qprop}] can be cast into the form
\begin{equation}\label{ad10a}
S(\vp) = \frac{\valpha\cdot\vp + \beta \, M_p}{2 \sqrt{p^2+\smash[b]{M^2_p}}}
\end{equation}
where the mass function $M_p$ is related to the variational kernel $s_p$ through
\begin{equation}\label{ad11}
M_p = \frac{2 p \, w_p + m ( 1-w^2_p )}{1 - w^2_p - 2 \frac{m}{p} \, w_p} \, .
\end{equation}
Equation \eqref{ad10a} gives a quasi-particle approximation to the full quark propagator:
It has the same form as the free-fermion propagator $S_0$ [\Eqref{evg4}] except that
the current quark mass $m$ is replaced by a running mass $M_p$. Note also that in this
case the quark dressing function becomes $Z_p=1$.

Equation \eqref{ad11} can be used to trade the kernel $s_p$ in the gap equation
\Eqref{ad8} for the running mass $M_p$ yielding
\[
M(\vp) = m + \frac{g^2 C_F}{2} \int \dfr[3]{q} \:
\frac{F(\vp-\vq)}{\sqrt{q^2+\smash[b]{M^2_q}}} \left[ M_q - \frac{\vp\cdot\vq}{\vp^2} \: M_p \right] .
\]
The same equation has been derived in Ref.~\cite{Watson:2011kv} from a truncated system
of DSEs in the so-called first order formalism.


\section{\label{sec:bare}The Bare-Vertex Approximation}

Let us now return to the full equations of motion of Sec.~\ref{sec:qcd} with the quark-gluon vertex included.
We are interested here mainly in recovering the results of Refs.~\cite{Vastag:2015qjd,Campagnari:2016wlt}
within the present CRDSE approach. For this purpose we replace in the following the full
quark-gluon vertex $\bar\Gamma$ [\Eqref{qgv}] by the bare one $\bar\Gamma_0$ [\Eqref{bqgv}].
We are aware that this approximation might not yet be entirely sufficient to provide a realistic
description of the mechanism of spontaneous breaking of chiral symmetry, i.e.~to yield
realistic values for quark condensate in agreement with low-energy meson phenomenology.
Nevertheless,
it is certainly worthwhile to investigate first the bare-vertex approximation in order
to get a better understanding of the structure of the equations of motion of the present
approach. In addition, the use of a bare quark-gluon vertex is sufficient to carry out
the renormalization of these equations, since the leading UV behaviour of the dressed
vertex agrees with that of the bare one, due to asymptotic freedom.

\subsection{The Quark CRDSE}

After replacing the full vertices in the CRDSE \eqref{1010-17a} by bare ones, the Dirac
traces can be worked out and the coupled equations \eqref{1010-17a} for the dressing
functions of the quark propagator reduce in the chiral limit $m=0$ to the set of equations~\eqref{qdse6}
given in Appendix~\ref{app:qdse}. Equations~\eqref{qdse6b} and \eqref{qdse6c} for the
dressing functions $B_p$ and $C_p$ can be collected into a single equation
for the complex quantity $H=B+\I* C$ 
\[
H_p = s_p + \frac{g^2 C_F}{2} \int \dfr[3]{q} \, \frac{H^*_q}{\Omega(\vp+\vq) \, \Delta_q}
\bigl[ X_-(\vp,\vq) \, V(\vp,\vq) \, V(\vq,\vp) - X_+(\vp,\vq) \, W(\vp,\vq) \, W(\vq,\vp) \bigr].
\]
where we have introduced the abbreviations
\begin{equation}\label{qdse7}
X_\pm(\vp,\vq) = 1 \pm \frac{[\uvp\cdot(\vp+\vq)] [\uvq\cdot(\vp+\vq)]}{(\vp+\vq)^2} \, ,
\end{equation}
while $\Delta_q$ is given by \Eqref{qdse8}.
Similarly, the equations \eqref{qdse6a} and \eqref{qdse6d} for $A_p$ and $D_p$ can be
added and subtracted, yielding
\begin{align*}
A_p+D_p &= 1 + \frac{g^2 C_F}{2} \int \dfr[3]{q} \: \frac{A_q+D_q}{\Omega(\vp+\vq) \, \Delta_q}
\biggl[ X_-(\vp,\vq) \, \abs{V(\vp,\vq)}^2 + X_+(\vp,\vq) \, \abs{W(\vp,\vq)}^2  \biggr] , \\
A_p-D_p &= \frac{g^2 C_F}{2} \int \dfr[3]{q} \, \frac{A_q-D_q}{\Omega(\vp+\vq) \, \Delta_q}
\biggl[ X_-(\vp,\vq) \, \abs{V(\vq,\vp)}^2 + X_+(\vp,\vq) \, \abs{W(\vq,\vp)}^2  \biggr] .
\end{align*}

If the variational vector kernels have the symmetry
\[
\abs{V(\vp,\vq)} = \abs{V(\vq,\vp)}, \qquad \abs{W(\vp,\vq)} = \abs{W(\vq,\vp)}
\]
then the Eqs.~\eqref{qdse6a} and \eqref{qdse6d} for the form factors $A_p$ and $D_p$
decouple and there exists always the trivial solution $D_p = 0$. Finally, notice
that for vanishing vector kernels $V=0=W$ these equations reduce to
\begin{equation}\label{g10}
A_p = 1 , \qquad B_p = \Re s_p , \qquad C_p = \Im s_p , \qquad D_p = 0 .
\end{equation}
The quark propagator is then entirely determined by the scalar variational kernel
$s_p$, which corresponds to the BCS-type model considered in Refs.~\cite{Adler:1984ri,Finger:1981gm,Alkofer:1988tc}, see Sect.~\ref{sec:AD}.
 

\subsection{Determination of the Variational Kernels}

From both continuum \cite{Watson:2011kv} as well as lattice \cite{Burgio:2012ph,Pak:2015dxa} studies
there exists no indication that the quark propagator in Coulomb gauge contains a term proportional to the Dirac
matrix $\beta\alpha_i$ [see \Eqref{qdse5a}]. Furthermore, when the energy variable of the full
propagator is integrated out to yield the
equal-time propagator, the term in the quark propagator proportional to the unit matrix vanishes
too. Therefore, we expect the physical quark propagator \Eqref{qdse5a} to be characterized by $C_p=D_p=0$.
It is not difficult to see that the quark CRDSEs~\eqref{qdse6} allow for consistent solutions
with $C_p=D_p=0$ when the variational kernels $s$, $V$ and $W$ are real, and the vector
kernels $V$ and $W$ are symmetric in the quark momenta. Under these assumptions the
quark propagator CRDSEs~\eqref{qdse6} reduce to
\begin{subequations}\label{loqcd1}
\begin{align}
\label{loqcd1a}
A_p &= 1 + g^2 \frac{C_F}{2} \int \dfr[3]{q} \, \frac{A_q}{\Delta_q \, \Omega(\vp+\vq)} \bigl[ X_-(\vp,\vq) \, V^2(\vp,\vq) + X_+(\vp,\vq) \, W^2(\vp,\vq) \bigr] , \\
B_p &= s_p + g^2 \frac{C_F}{2} \int \dfr[3]{q} \, \frac{B_q}{\Delta_q \, \Omega(\vp+\vq)} \bigl[ X_-(\vp,\vq) \, V^2(\vp,\vq) - X_+(\vp,\vq) \, W^2(\vp,\vq) \bigr] ,
\end{align}
\end{subequations}
while the contributions to the energy density [see Eqs.~\eqref{eD2}, \eqref{eE7}, and \eqref{eCqq6}] become
\begin{subequations}\label{loe1}
\begin{align}
\label{loe1a}
e^{}_\mathrm{D} &= 
- 4 \int\dfr[3]{q} \, \frac{\abs{\vq} \, A_q}{\Delta_q} \nonumber \\
&{}\qquad + 2 g^2 C_F \int \dfr[3]{q} \dfr[3]{\ell} \, 
\frac{X_-(\vq,\vl) \, V(\vq,\vl)(A_q \, A_\ell + B_q \, B_\ell) + X_+(\vq,\vl) \, W(\vq,\vl)(A_q \, B_\ell + B_q \, A_\ell)}{\Delta_q \, \Delta_\ell \, \Omega(\vq+\vl)} , \\
\label{loe1b}
e_E^\mathrm{Q} &= g^2 C_F \int \dfr[3]{q} \dfr[3]{\ell} \, \frac{A_q A_\ell}{\Delta_q \Delta_l} \bigl[ X_-(\vq,\vl) \, V^2(\vq,\vl) + X_+(\vq,\vl) \, W^2(\vq,\vl) \bigr] , \\
\label{loe1c}
e_\mathrm{C}^{qq} &= - g^2 \frac{C_F}{2} \int \dfr[3]{q} \dfr[3]{\ell} \: F(\vq-\vl)
\frac{4 B_q \, B_\ell + \uvq\cdot\uvl \bigl[ A_q(2-A_q)-B_q^2\bigr] \bigl[ A_\ell(2-A_\ell)-
B_\ell^2\bigr]}{\Delta_q \Delta_\ell} ,
\end{align}
\end{subequations}
with $\Delta$ [\Eqref{qdse8}] reducing to
\begin{equation}\label{nug11}
\Delta_p = A^2_p + B^2_p .
\end{equation}
The energy density contributions \eqref{loe1} contain the scalar kernel $s_p$ only
implicitly through the dressing functions $A_p$ and $B_p$, while the vector kernels $V$ and
$W$ enter both explicitly and implicitly. From \Eqref{loqcd1a} we find the derivatives
of the dressing function $A_p$ with respect to the vector kernels
\begin{align}
\label{loe2}
\frac{\delta A_k}{\delta V(\vp,\vq)} ={}&
g^2 \frac{C_F}{2} \frac{X_-(\vp,\vq)}{\Omega(\vp+\vq)}
V(\vp,\vq) \left[\deltabar(\vk-\vp) \frac{A_q}{\Delta_q} + \deltabar(\vk-\vq) \frac{A_p}{\Delta_p} \right] + \dots \\
\label{loe2w}
\frac{\delta A_k}{\delta W(\vp,\vq)} ={}&
g^2 \frac{C_F}{2} \frac{X_+(\vp,\vq)}{\Omega(\vp+\vq)}
W(\vp,\vq) \left[\deltabar(\vk-\vp) \frac{A_q}{\Delta_q} + \deltabar(\vk-\vq) \frac{A_p}{\Delta_p} \right]+ \dots
\end{align}
and similarly the derivatives of $B_p$
\begin{align}
\label{loe2nv}
\frac{\delta B_k}{\delta V(\vp,\vq)} &=
g^2 \frac{C_F}{2} \frac{X_-(\vp,\vq)}{\Omega(\vp+\vq)}
V(\vp,\vq) \left[\deltabar(\vk-\vp) \frac{B_q}{\Delta_q} + \deltabar(\vk-\vq) \frac{B_p}{\Delta_p} \right] + \dots \\
\label{loe2nw}
\frac{\delta B_k}{\delta W(\vp,\vq)} &= -
g^2 \frac{C_F}{2} \frac{X_+(\vp,\vq)}{\Omega(\vp+\vq)}
W(\vp,\vq) \left[\deltabar(\vk-\vp) \frac{B_q}{\Delta_q} + \deltabar(\vk-\vq) \frac{B_p}{\Delta_p} \right] + \dots
\end{align}
The ellipsis on the right-hand side of these equations stand for the one-loop terms,
which we will usually neglect since they would give rise to more than one loop in the
equations of motion of the vector kernels.

In the same way we can evaluate the functional derivatives of the dressing functions $A_p$
and $B_p$ with respect to the scalar kernel~$s_p$
\begin{align*}
\frac{\delta A_k}{\delta s_p} &=
- g^2 \frac{C_F}{2} \int \dfr[3]{q}\frac{X_-(\vk,\vq) \, V^2(\vk,\vq) + X_+(\vk,\vq) \, W^2(\vk,\vq)}{\Delta_q^2 \, \Omega(\vk+\vq)}
\biggl[ \bigl( A^2_q-B^2_q \bigr) \frac{\delta A_q}{\delta s_p} + 2 A_q B_q \frac{\delta B_q}{\delta s_p} \biggr] ,
\\
\frac{\delta B_k}{\delta s_p} &= \deltabar(\vp-\vk)
+ g^2 \frac{C_F}{2} \int \dfr[3]{q}\frac{X_-(\vk,\vq) \, V^2(\vk,\vq) - X_+(\vk,\vq) \, W^2(\vk,\vq)}{\Delta_q^2 \, \Omega(\vk+\vq)}
\biggl[ \bigl( A^2_q-B^2_q \bigr) \frac{\delta B_q}{\delta s_p} - 2 A_q B_q \frac{\delta A_q}{\delta s_p} \biggr] .
\end{align*}
At one-loop order the previous equations reduce to
\begin{equation}\label{vds}
\begin{split}
\frac{\delta A_k}{\delta s_p} &= - g^2 C_F \frac{A_p B_p}{\Delta_p^2 \, \Omega(\vk+\vp)} \bigl[ X_-(\vk,\vp) \, V^2(\vk,\vp) + X_+(\vk,\vp) \, W^2(\vk,\vp)\bigr] + \dots \\
\frac{\delta B_k}{\delta s_p} &= \deltabar(\vp-\vk)
+ g^2 \frac{C_F}{2} \frac{A_p^2-B_p^2}{\Delta_p^2 \, \Omega(\vk+\vp)} \bigl[ X_-(\vk,\vp) \, V^2(\vk,\vp) - X_+(\vk,\vp) \, W^2(\vk,\vp)\bigr] + \dots
\end{split}
\end{equation}

In a diagrammatic language, differentiating with respect to the vector kernel implies
removing one quark-gluon vertex from the diagram. Since the energy contributions contain
at most two loops, the variational equations for $V$ and $W$ are free of loops.
To this order, we can ignore the
Coulomb energy \Eqref{loe1c} and include only the explicit dependence on $V$ and $W$ in the
second term of \Eqref{loe1a} and in \Eqref{loe1b}, yielding
\begin{equation}\label{vdv1}
\frac{\delta \bigl(e_\mathrm{D}^{(1)}+e_E^q\bigr)}{\delta V(\vp,\vq)}
= 2 g^2 C_F \frac{X_-(\vp,\vq)}{\Delta_p \Delta_q} \biggl[ \frac{A_p A_q + B_p B_q}{\Omega(\vp+\vq)} + A_p A_q V(\vp,\vq) \biggr]
\end{equation}
as well as
\[
\frac{\delta \bigl(e_\mathrm{D}^{(1)}+e_E^q\bigr)}{\delta W(\vp,\vq)}
= 2 g^2 C_F \frac{X_+(\vp,\vq)}{\Delta_p \Delta_q} \biggl[ \frac{A_p B_q + B_p A_q}{\Omega(\vp+\vq)} + A_p A_q W(\vp,\vq) \biggr] .
\]
In the first term of \Eqref{loe1a}, however, we must take into account also the dependence
of the dressing functions $A_p$ and $B_p$ on the kernels $V$ and $W$. This yields
\[
\frac{\delta e_\mathrm{D}^{(0)}}{\delta V(\vp,\vq)} = 4 \int\dfr[3]\ell \, \frac{\abs{\vl}}{\Delta_\ell^2}
\biggl\{ (A_\ell^2-B_\ell^2) \frac{\delta A_\ell}{\delta V(\vp,\vq)} + 2 A_\ell B_\ell \, \frac{\delta B_\ell}{\delta V(\vp,\vq)} \biggr\}
\]
and by using Eqs.~\eqref{loe2} and \eqref{loe2nv} we find
\begin{equation}\label{vdv2}
\frac{\delta e_\mathrm{D}^{(0)}}{\delta V(\vp,\vq)} = 2 g^2 C_F \frac{X_-(\vp,\vq)}{\Omega(\vp+\vq)} \, V(\vp,\vq)
\biggl\{ \frac{\abs{\vp}}{\Delta_p^2} \biggl[ (A_p^2-B_p^2) \frac{A_q}{\Delta_q} + 2 A_p B_p \frac{B_q}{\Delta_q} \biggr]
+ \frac{\abs{\vq}}{\Delta_q^2} \biggl[ (A_q^2-B_q^2) \frac{A_p}{\Delta_p} + 2 A_q B_q \frac{B_p}{\Delta_p} \biggr] \biggr\} .
\end{equation}
Requiring that the sum of Eqs.~\eqref{vdv1} and \eqref{vdv2} vanishes fixes the vector
kernel $V$ to
\begin{equation}\label{loqcd5}
V(\vp,\vq) =
- \frac{A_p A_q + B_p B_q}{A_p A_q\Omega(\vp+\vq) + \abs{\vp} \frac{A_q(A_p^2-B_p^2)+2A_p B_p B_q}{\Delta_p} + \abs{\vq} \frac{A_p(A_q^2-B_q^2)+2A_q B_p B_q}{\Delta_p}} .
\end{equation}
To simplify this and the following expressions we introduce the ratio
\begin{equation}\label{loqcd3}
b_p \equiv \frac{B_p}{A_p}
\end{equation}
and cast \Eqref{loqcd5} into the form
\begin{equation}\label{loqcd4}
V(\vp,\vq) = - \frac{1 + b_p b_q}{\Omega(\vp+\vq) + \frac{\abs{\vp}}{A_p} \frac{1-b_p^2+2 b_p b_q}{1+b_p^2} + \frac{\abs{\vq}}{A_q} \frac{1-b_q^2+2b_p b_q}{1+b_q^2}} .
\end{equation}
At leading order we find from \Eqref{loqcd1} $A_p=1$ and $b_p=s_p$, and \Eqref{loqcd4} reduces to the kernel
found in Ref.~\cite{Vastag:2015qjd}. Furthermore, at large momenta we recover
the leading-order perturbative result \cite{Campagnari:2014hda}.

The variation of the energy with respect to~$W$ is carried out in an analogous way
by using Eqs.~\eqref{loe2w} and \eqref{loe2nw}. This yields the equation of motion
\begin{equation}
\label{loqcd4w}
W(\vp,\vq) = - \frac{b_p + b_q}{\Omega(\vp+\vq) + \frac{\abs{\vp}}{A_p} \frac{1-b_p^2-2 b_p b_q}{1+b_p^2} + \frac{\abs{\vq}}{A_q} \frac{1-b_q^2-2b_p b_q}{1+b_q^2}} .
\end{equation}
Also this kernel reduces to the one found in Ref.~\cite{Vastag:2015qjd} at leading order.
Both kernels $V$ and $W$ turn out to be real and negative, as we might have expected from
$e_{\mathrm{D}}^{(1)}$ [\Eqref{lve1bis}]: this is the only energy contribution involving
the variational vector kernels linearly. This energy contribution vanishes if the
quark-gluon coupling is neglected in the vacuum wave functional, i.e.~for $V=0=W$. 
Negative vector kernels $V$ and $W$ are energetically favoured since they make $e_\mathrm{D}^{(1)}$
negative.

The variation of the energy density with respect to the scalar kernel~$s_p$ is slightly
more involved than the variational derivative with respect to the vector kernels.
For the second term in the single-particle energy density \Eqref{loe1a}, as well as for the
contributions of the gluonic kinetic term \Eqref{loe1b} and of the Coulomb interaction \Eqref{loe1c}
it is sufficient to keep only the leading order of \Eqref{vds}, while for the first term in
\Eqref{loe1a} we need also the one-loop contributions.
Then the variation with respect to $s_p$ yields
\begin{equation}\label{nge1}
\begin{aligned}[b]
\frac{b_p \abs{\vp}}{A_p^2(1+b_p^2)^2} ={}& \frac{g^2 C_F}{2 A_p^2(1+b_p^2)^2} \int\dfr[3]{q} \, \frac{1}{A_q (1+b_q^2)}
\biggl\{
 b_p \bigl[ X_-(\vp,\vq) \, V^2(\vp,\vq) + X_+(\vp,\vq) \, W^2(\vp,\vq) \bigr] \\
& - \frac{\abs{\vq}}{A_q (1+b_q^2) \Omega(\vp+\vq)}
  \begin{aligned}[t]
\Bigl[&
  X_-(\vp,\vq) \, V^2(\vp,\vq) \bigl[ (1-b_p^2)b_q - b_p(1-b_q^2) \bigr] \\
  {}&- X_+(\vp,\vq) \, W^2(\vp,\vq) \bigl[(1-b_p^2)b_q+b_p(1-b_q^2) \bigr]
\Bigr]
\end{aligned}
\\
&-\frac{1}{\Omega(\vp+\vq)}
\Bigl[
  X_-(\vp,\vq) \, V(\vp,\vq) \bigl[ (1-b_p^2)b_q - 2b_p) \bigr]
  + X_+(\vp,\vq) \, W(\vp,\vq) \bigl[1-b_p^2-2b_pb_q \bigr]
\Bigr] \\
&+ F(\vp-\vq) \Bigl[ b_q (1-b_p^2) - \uvp\cdot\uvq \bigl(2-A_q(1+b_p^2)\bigr) \Bigr] \biggr\} ,
\end{aligned}
\end{equation}
where we have expressed the resulting equations in terms of $b_p$ \Eqref{loqcd3}
instead of $s_p$.
In order to reproduce the loop expansion of Ref.~\cite{Vastag:2015qjd}\footnote{%
   The present approach allows one to go beyond this loop expansion.}
on the right-hand side of \Eqref{nge1} it is sufficient to replace $b_p\to s_p$ and $A_p\to1$,
while on the left-hand side the factor
\[
\frac{b_p}{A_p^2(1+b_p^2)^2} = \frac{A_p B_p}{(A_p^2+B_p^2)^2}
\]
has to be expanded up to one-loop order by means of Eqs.~\eqref{loqcd1}, yielding
\begin{multline*}
\frac{b_p}{A_p^2(1+b_p^2)^2}
= \frac{s_p}{(1+s_p^2)^2} + \frac{1}{(1+s_p^2)^3} \frac{g^2C_F}{2}
\int\dfr[3]{q} \frac{1}{(1+s_q^2) \Omega(\vp+\vq)} \\
\times \Bigl\{ X_-(\vp,\vq) \, V^2(\vp,\vq) \bigl[ s_p(s_p^2-3)+s_q(1-3s_p^2) \bigr]
+ X_+(\vp,\vq) \, W^2(\vp,\vq)\bigl[ s_p(s_p^2-3) - s_q(1-3s_p^2) \bigr] \Bigr\} .
\end{multline*}
With these replacements \Eqref{nge1} reduces precisely to the gap equation found in
Refs.~\cite{Vastag:2015qjd,Campagnari:2016wlt}, which is explicitly given in our notation
in Appendix~\ref{app:gap}.
In fact, the same result may be obtained by
expanding the dressing functions \eqref{loqcd1} at one-loop order in the
energy density contributions \eqref{loe1} and taking the variation afterwards.


\section{\label{sec:app}Renormalized Quark Propagator and Chiral Condensate}

The gap equation \eqref{x1} has been solved numerically for the variational kernel $s_p$
in Ref.~\cite{Campagnari:2016wlt}. The renormalization of the quark propagator \Eqref{prop3}
was ignored and the quark condensate was evaluated from the leading-order (in the number
of quark loops) propagator
\begin{equation}\label{bp}
S(\vp) = \frac{(1-s_p^2) \valpha\cdot\uvp + 2 s_p \beta}{2(1+s_p^2)},
\end{equation}
which arises from the full propagator \Eqref{prop3} by putting $A_p=1$ and $B_p=s_p$,
which are the zero-loop expressions [see \Eqref{g10}].
The coupling $g$ was then chosen to reproduce the phenomenological value of the quark
condensate
\[
\vev{\bar{q}q} = - \int\dfr[3]{p} \tr\bigl[ \beta S(\vp) \bigr] = - \frac{2\Nc}{\pi^2} \int \d p \, p^2 \, \frac{s_p}{1+s_p^2} .
\]

Here we go beyond Ref.~\cite{Campagnari:2016wlt} and consistently calculate the quark
propagator up to including one-loop order. This should be sufficient to investigate the
renormalization properties of the quark propagator.

To one-loop order we can replace the denominator $\Delta_p$ [\Eqref{nug11}]
in \Eqref{loqcd1} by its leading-order expression $A_p=1$, $B_p=s_p$.
Then Eqs.~\eqref{loqcd1} become
\[
\begin{split}
A_p &= 1 + g^2 \frac{C_F}{2} \int \dfr[3]{q} \, \frac{X_-(\vp,\vq) \, V^2(\vp,\vq) + X_+(\vp,\vq) \, W^2(\vp,\vq)}{(1+s_q^2) \, \Omega(\vp+\vq)}
       \equiv 1 + I_A(p,\Lambda) ,\\
B_p &= s_p + g^2 \frac{C_F}{2} \int \dfr[3]{q} \, s_q \, \frac{X_-(\vp,\vq) \, V^2(\vp,\vq) - X_+(\vp,\vq) \, W^2(\vp,\vq)}{(1+s_q^2) \, \Omega(\vp+\vq)}
       \equiv s_p + I_B(p,\Lambda) ,
\end{split}
\]
where $V$ and $W$ are given by Eqs.~\eqref{loqcd4} and \eqref{loqcd4w} with $b_p$ replaced
by $s_p$ and $A_p$ replaced by 1; furthermore, $\Lambda$ is a momentum cut-off. A quick calculation
shows that the loop integral~$I_B$ is convergent while $I_A$ is logarithmically divergent
\begin{equation}\label{ren1}
I_A(p,\Lambda) = \frac{g^2 C_F}{(4\pi)^2} (1+s_p^2) \ln\Lambda + \text{finite terms}.
\end{equation}
At first sight, the appearance of a momentum-dependent divergence seems to spoil
multiplicative renormalizability. However, this is not the case, as we will show now.
Expanding the quark propagator \Eqref{prop3} at one-loop order we obtain
\[
S(\vp,\Lambda) = \frac{1}{1+s_p^2} \biggl\{
\frac{\valpha\cdot\uvp}{2} \biggl[ (1-s_p^2)\biggl(1-\frac{2 I_A(p,\Lambda)}{1+s_p^2}\biggr) - \frac{4 s_p I_B(p,\Lambda)}{1+s_p^2} \biggr]
+ \beta \biggl[ s_p\biggl(1-\frac{2 I_A(p,\Lambda)}{1+s_p^2}\biggr) + \frac{1-s_p^2}{1+s_p^2} \, I_B(p,\Lambda) \biggr] \biggr\} .
\]
Inserting here \Eqref{ren1} one finds that the momentum-dependent part of the logarithmic
divergence cancels. The remaining part of the UV divergence can be removed by
the perturbative one-loop
renormalization constant, which in the $\overline{\text{MS}}$ scheme reads
\cite{Popovici:2008ty,Campagnari:2014hda}
\begin{equation}\label{ren2}
Z_2(\Lambda,\mu) = 1 - \, \frac{g^2 C_F}{(4\pi)^2} \biggl[ \ln\frac{\Lambda^2}{\mu^2} + \ln4\pi - \gamma^{}_\mathrm{E} \biggr] \equiv 1 - \delta z_2(\Lambda,\mu) .
\end{equation}
With this expression we can define a renormalized propagator
\begin{equation}\label{ren19}
S(\vp,\mu) = f_\alpha(p,\mu) \, \valpha\cdot\uvp + f_\beta(p,\mu) \, \beta 
\end{equation}
where
\begin{equation}\label{ren3}
\begin{split}
f_\alpha(p,\mu) &= \frac12 \frac{1-s_p^2}{1+s_p^2} \biggl[ 1-\frac{2 I_A(p,\Lambda)}{1+s_p^2} + \delta z_2(\Lambda,\mu) \biggr] - \frac{2 s_p}{(1+s_p^2)^2} \, I_B(p) , \\
f_\beta(p,\mu) &= \frac{s_p}{1+s_p^2} \biggl[ 1-\frac{2 I_A(p,\Lambda)}{1+s_p^2} + \delta z_2(\Lambda,\mu) \biggr] + \frac{1-s_p^2}{(1+s_p^2)^2} \, I_B(p) .
\end{split}
\end{equation}
By means of Eqs.~\eqref{ren1} and \eqref{ren2} one finds that \Eqref{ren3} is indeed
finite when the cut-off $\Lambda$ is removed.

It is important to note that without the vector kernel $W$, i.e.~without the Dirac
structure $\beta \alpha_i$ in the bare quark-gluon vertex of the quark wave functional,
the term $1+s_p^2$ in \Eqref{ren1} would reduce to $1$ and the physical quark propagator
would no longer be multiplicatively renormalizable [cf.~\Eqref{ren3}].
Although the vector kernel $W$ [\Eqref{loqcd4w}] is purely non-perturbative in nature,
its presence in the quark wave functional [Eqs.~\eqref{evg2}, \eqref{ans2}, \eqref{ans6}]
is necessary to ensure multiplicative renormalizability of the propagator.

\begin{figure}
\centering
\includegraphics[width=.4\linewidth]{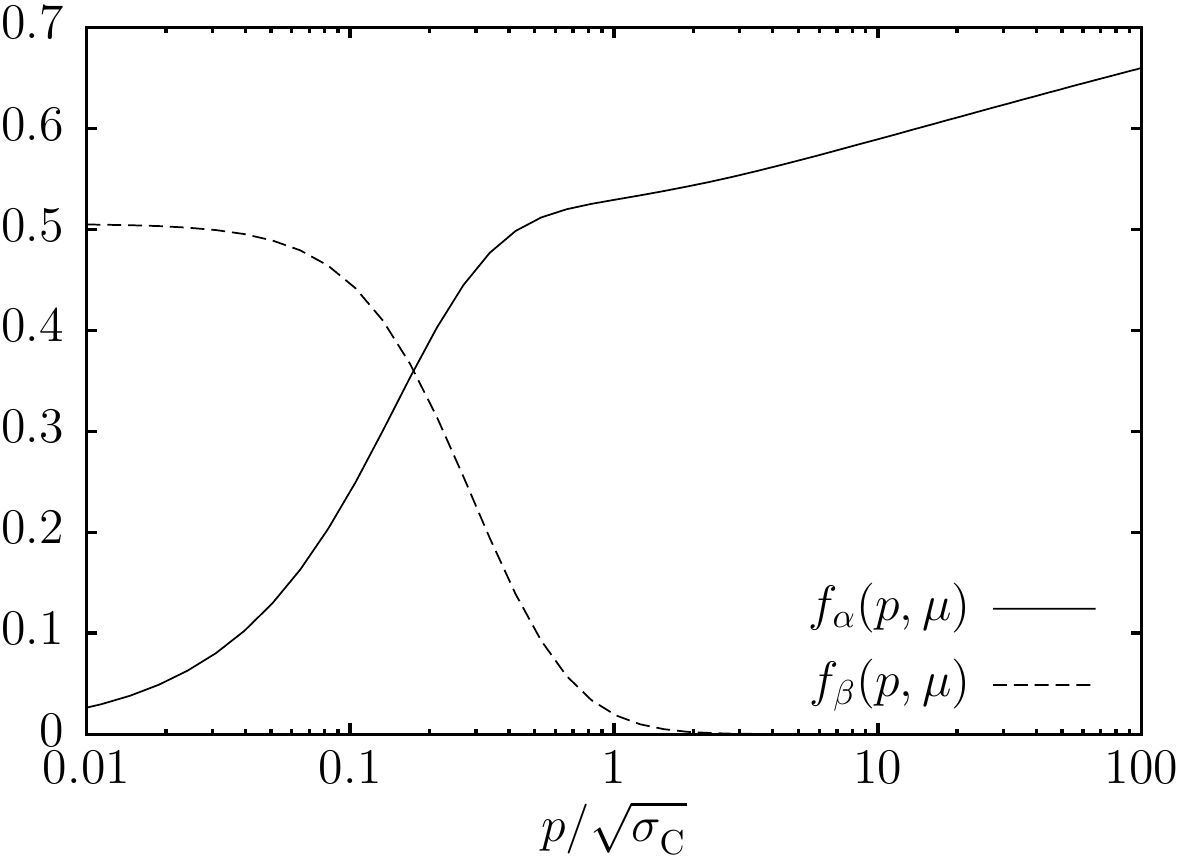}\qquad
\includegraphics[width=.4\linewidth]{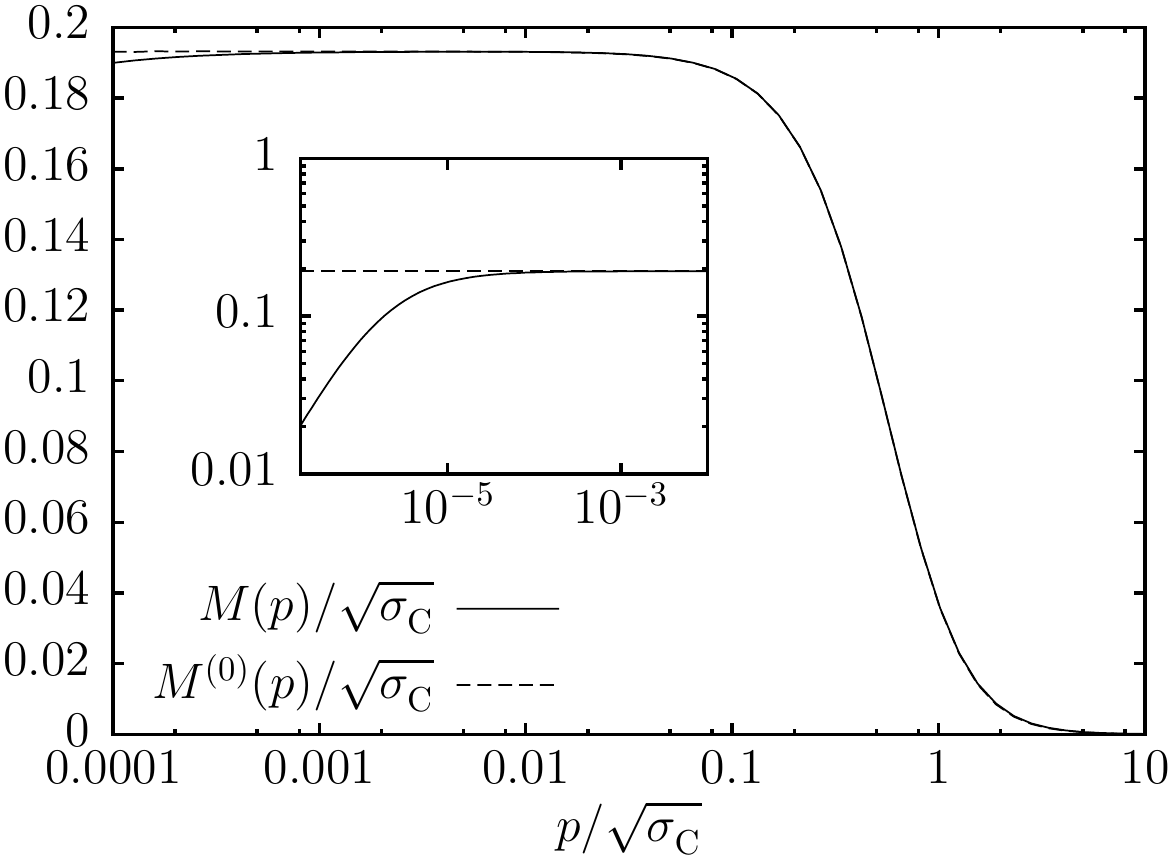}
\caption{(left) Dressing functions $f_\alpha$ and $f_\beta$ of the renormalized quark propagator.
(right) Mass function of the renormalized [\Eqref{ren4}, continuous line] and unrenormalized
[\Eqref{ren18}, dashed line] quark propagator.}
\label{fig:dressingfunctions}
\end{figure}

The renormalization point dependent quark condensate
in the $\overline{\text{MS}}$ scheme is usually quoted at the renormalization scale
$\mu=2\,\mathrm{GeV}$. At this scale the running strong coupling constant has the value
$\alpha_s(2\,\mathrm{GeV})=0.30(1)$ \cite{Buckley:2014ana}. Solving the gap equation~\eqref{x1}
with this value of $\alpha_s$ yields the dressing functions [\Eqref{ren3}] shown
in Fig.~\ref{fig:dressingfunctions}.
The resulting chiral condensate is
\[
\vev{\bar{q}q} \simeq \bigl(-0.31 \sqrt{\smash[b]{\sigma_\mathrm{C}^{}}} \bigr)^3 .
\]
The scale in our calculations is fixed by the Coulomb string tension $\sigma_\mathrm{C}$
occurring in the colour Coulomb potential \Eqref{coulpot}.
Lattice and continuum calculations \cite{Nakagawa:2006fk,Golterman:2012dx,Greensite:2014bua}
quote values of the Coulomb string tension from 2.5 to as large as 4 times the Wilson
string tension $\sigma=(440\,\mathrm{MeV})^2$, which gives us $\sqrt{\smash[b]{\sigma_\mathrm{C}^{}}}$
in the range from 696\,MeV to 880\,MeV. This yields a (renormalization point dependent)
chiral condensate in the range between $(-216\,\mathrm{MeV})^3$ and $(-270\,\mathrm{MeV})^3$.
Lattice simulations and chiral perturbation theory calculations yield for the chiral
condensate values in a similar range \cite{Borsanyi:2012zv,Cossu:2016eqs,Aoki:2016frl,Faber:2017alm,Tomii:2017cbt}.

The renormalized quark propagator \Eqref{ren19} can be cast into the form \Eqref{prop4}
\begin{equation}\label{g14}
S(\vp,\mu) = Z_p \, \frac{\valpha\cdot\vp + \beta M_p}{2\sqrt{\vp^2+M_p^2}}
\end{equation}
where the mass function $M_p$ and the dressing function $Z_p$ are related to $f_\alpha$ and
$f_\beta$ [\Eqref{ren3}] by
\begin{equation}\label{ren4}
M_p = \frac{\abs{\vp} \, f_\beta(p,\mu)}{f_\alpha(p,\mu)} , \qquad
Z_p = 2 \sqrt{f_\alpha^2(p,\mu)+f_\beta^2(p,\mu)} .
\end{equation}
From the definition of the mass function [\Eqref{ren4}] it is clear that if $f_\alpha$
does not vanish for $\vp=0$ the mass function is bound to vanish in the deep infrared.
Our numerical results show that while $s(0)=1$, $I_B(0)$ is very small but not vanishing.
The reason for this behaviour is the fact that the denominators of
the vector kernels $V$ [\Eqref{loqcd4}] and $W$ [\Eqref{loqcd4w}] are not the same. We believe
that this is an artefact of the one-loop expansion.
The mass function \Eqref{ren4} stays however
constant over almost three orders of magnitude before slowly bending over (see Fig.~\ref{fig:dressingfunctions}). Furthermore,
the integral $I_B(p)$ is rather small in comparison to the (renormalized) integral $I_A$.
While the latter has an important effect on the chiral condensate, the mass function \Eqref{ren4} is, apart
from the deep IR, almost indistinguishable from the mass function of Ref.~\cite{Campagnari:2016wlt}
extracted from the unrenormalized quark propagator \Eqref{bp}
\begin{equation}\label{ren18}
M^{(0)}(p) = \frac{2 p s_p}{1-s_p^2}
\end{equation}
as shown in Fig.~\ref{fig:dressingfunctions}.
While our mass function vanishes in the deep infrared, the plateau value reads
\[
M_\mathrm{IR} \simeq 0.19\,\sqrt{\smash[b]{\sigma_\mathrm{C}^{}}} ,
\]
which, due to the uncertainty in the Coulomb string tension, is in the range between 135
and 170 MeV.


\section{\label{sec:m3m4}Mass Function in the Full and Static Propagator}

As mentioned before,
in Ref.~\cite{Campagnari:2016wlt} the renormalization of the propagator was ignored and
the value of the quark-gluon coupling constant was chosen to reproduce the phenomenological value of
the chiral condensate. The mass function, however, was not significantly enhanced in
comparison to the Adler--Davis model \cite{Adler:1984ri} (see Sec.~\ref{sec:AD}), showing an infrared value of 135\,MeV
(for $\sigma_\mathrm{C}=2.5\sigma$). Similar results have been obtained also in the previous
section: although our rough one-loop calculation is capable of reproducing the correct
value of the chiral condensate, the mass function is not significantly influenced by the
coupling to the transverse gluons. This seems
at odds with the common lore that the infrared value of the mass function should be around
the value of the constituent quark mass, i.e.~roughly 300\,MeV. Here we show that this
apparent contradiction might result from comparing the mass functions of the full and equal-time
propagators. Before discussing this issue in Coulomb gauge we address the question in Landau
gauge, for which we have solutions of the Dyson--Schwinger equations at our disposal.

\begin{figure}
\parbox{.45\linewidth}{\centering\includegraphics[width=\linewidth]{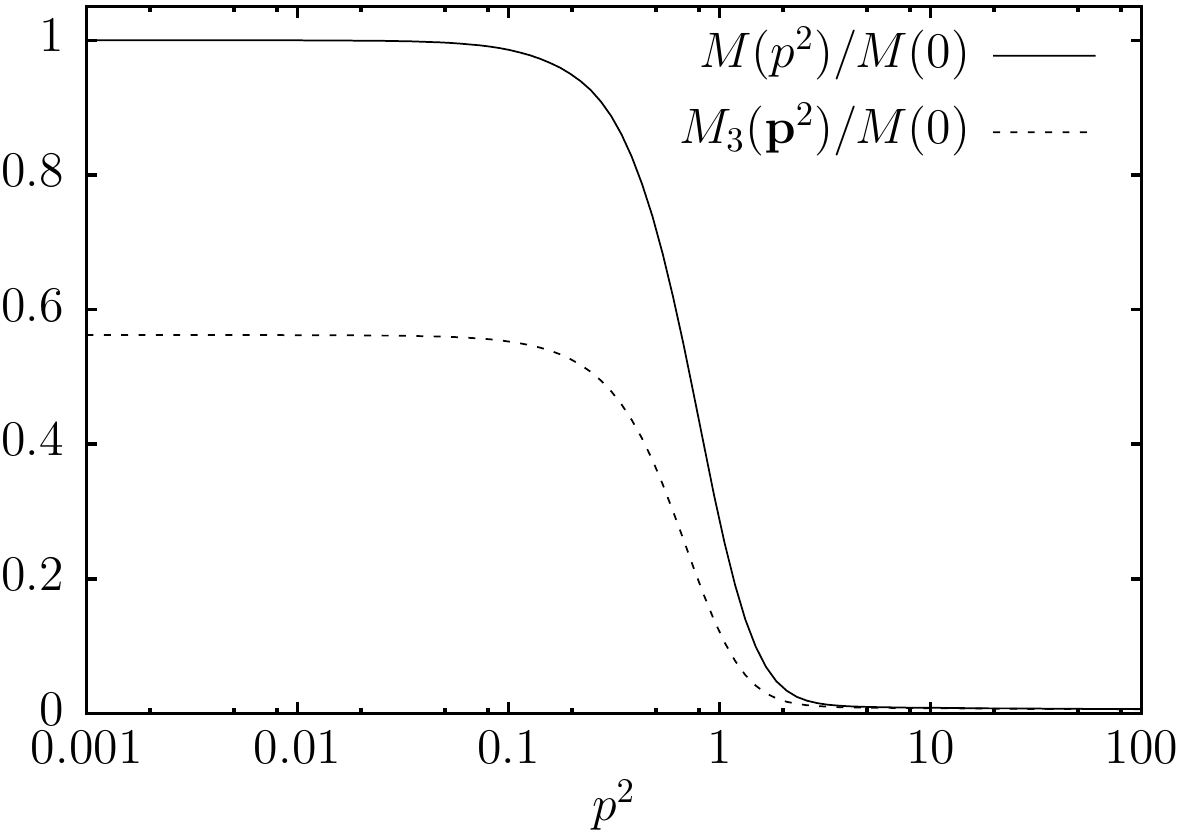}\\(a)}
\hfill
\parbox{.45\linewidth}{\centering\includegraphics[width=\linewidth]{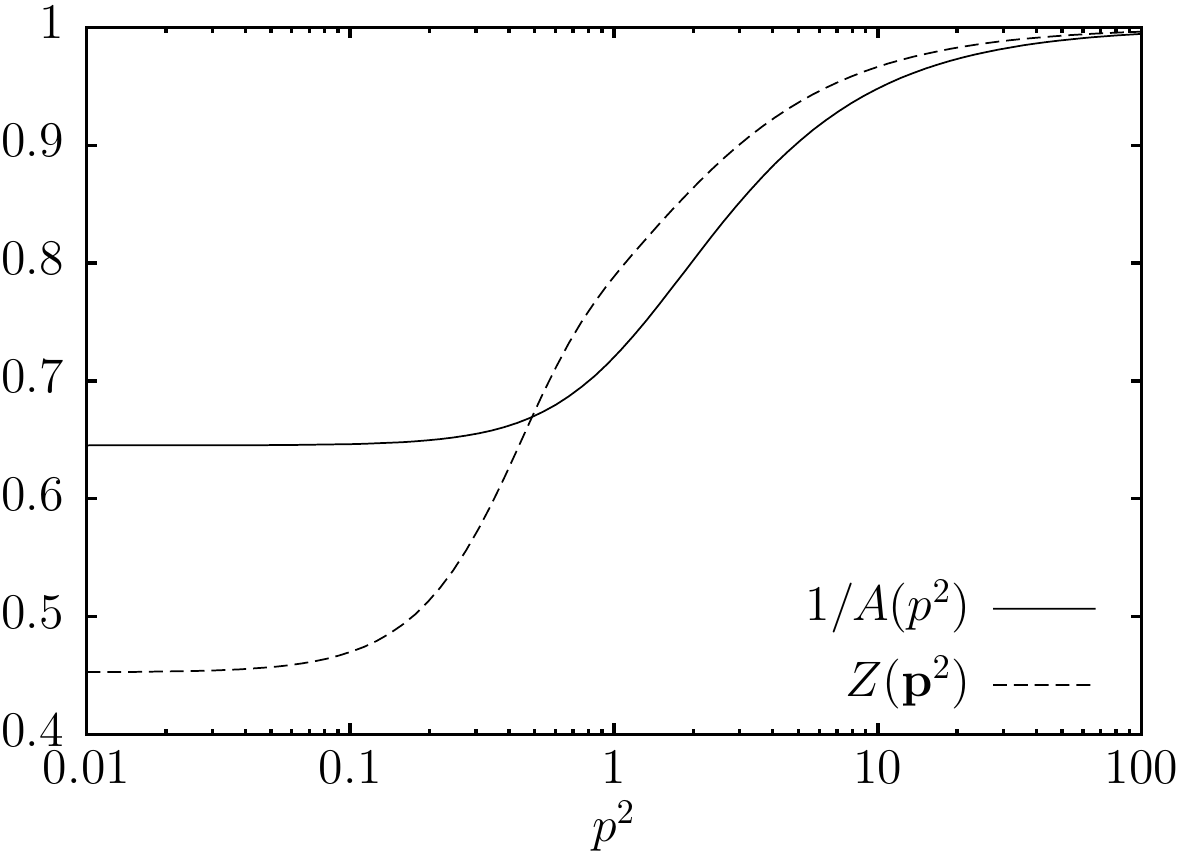}\\(b)}
\caption{\label{fig:staticmassfunction} (a) Comparison between the full mass function
$M(p^2)$ in Landau gauge (continuous line) and the mass function $M_3(\vp^2)$ of the
equal-time propagator (dashed line). (b) Dressing function of the full (continuous line)
and equal-time propagator (dashed line).}
\end{figure}

Suppressing colour indices, the quark propagator in Landau gauge is usually written as
\begin{equation}\label{g15}
S(p) = \frac{1}{-\I* \slashed{p} A(p^2) + B(p^2)}
     = \frac{1}{A(p^2)} \frac{\I* \slashed{p} + M(p^2)}{p^2 + M^2(p^2)} ,
\end{equation}
where the quark mass function $M$ is defined as $M(p^2) = B(p^2)/A(p^2)$. At tree level
we have $A=1$ and $B=M=m$, with $m$ being the bare current quark mass. The equal-time
propagator $S_3(\vp)$ is obtained from the full one $S(p)$ by integrating out the energy component $p_4$
of the four-momentum
\[
S_3(\vp) = \int\dfr{p_4} \, S(p) .
\]
For symmetry reasons the contribution proportional to $\gamma_4 p_4$ vanishes and we are left with
\begin{equation}\label{x2}
S_3(\vp) = \I* \vec\gamma\cdot\vp \int\dfr{p_4} \frac{1}{A(p_4^2+\vp^2)} \frac{1}{p_4^2+\vp^2 + M^2(p_4^2+\vp^2)}
+ \int\dfr{p_4} \frac{1}{A(p_4^2+\vp^2)} \frac{M(p_4^2+\vp^2)}{p_4^2+\vp^2 + M^2(p_4^2+\vp^2)} .
\end{equation}
Analogously to the definition of the quark mass function $M$ we can introduce the
equal-time mass function $M_3(\vp^2)$ as ratio of the coefficients of the $\id$ and $\gamma^i$
terms of the equal-time propagator, yielding
\begin{equation}\label{lgetm1}
M_3(\vp^2) =
\frac{\displaystyle\int_0^\infty \d p_4 \, \frac{1}{A(p_4^2+\vp^2)} \frac{M(p_4^2+\vp^2)}{p_4^2+\vp^2 + M^2(p_4^2+\vp^2)}}
     {\displaystyle\int_0^\infty \d p_4 \, \frac{1}{A(p_4^2+\vp^2)} \frac{1}{p_4^2+\vp^2 + M^2(p_4^2+\vp^2)}}
\end{equation}
Numerical solutions for the mass function always show a monotonically decreasing function
of the four-momentum. Therefore, since $M(p^2)\leq M(0)$ we see from \Eqref{lgetm1} that
$M_3(0) < M(0)$. For typical results for the Landau gauge quark propagator we
find that $M_3(0)$ lies between 50\% and 60\% of $M(0)$, see Fig.~\ref{fig:staticmassfunction}a.
Furthermore, the equal-time quark propagator \Eqref{x2} can be brought into the form \eqref{g14}
\[
S_3(\vp) = Z(\vp^2) \frac{\I*\vec\gamma\cdot\vp + M_3(\vp^2)}{2 \sqrt{\vp^2 + M_3^2(\vp^2)}} .
\]
Figure \ref{fig:staticmassfunction}b shows both $Z$ and $A^{-1}$.

The situation might be similar in Coulomb gauge. Being non-covariant, the propagator depends
\emph{separately} on $p_4$ and $\vp$ and has therefore \emph{four} Dirac components instead of two
\[
S^{-1}(p) = -\I* \gamma_4 p_4 A_t(p_4,\vp) - \I* \vec\gamma \cdot \vp A_s(p_4,\vp) - \I* \gamma_4 p_4 \vec\gamma \cdot\vp A_d(p_4,\vp) + B(p_4,\vp) .
\]
The mixed structure $\gamma_4 \gamma_i$ does not arise at one-loop level in perturbation theory \cite{Popovici:2008ty}
and is not found in lattice calculations \cite{Burgio:2012ph,Pak:2015dxa} either; therefore we
will set $A_d=0$ in the following. The propagator in Coulomb gauge takes therefore
the form
\[
S(p) = \frac{\I* \gamma_4 p_4 A_t(p_4,\vp) + \I* \vec\gamma \cdot \vp A_s(p_4,\vp) + B(p_4,\vp)}{p_4^2 A_t^2(p_4,\vp) + \vp^2 A_s^2(p_4,\vp) + B^2(p_4,\vp)} .
\]
Analogously to \Eqref{lgetm1} the equal-time mass function in Coulomb gauge is given by
\[
M_3(\vp) =
\frac{\displaystyle\int_0^\infty \d p_4 \, \frac{B(p_4,\vp)}{p_4^2 A_t^2(p_4,\vp) + \vp^2 A_s^2(p_4,\vp) + B^2(p_4,\vp)}}
     {\displaystyle\int_0^\infty \d p_4 \, \frac{A_s(p_4,\vp)}{p_4^2 A_t^2(p_4,\vp) + \vp^2 A_s^2(p_4,\vp) + B^2(p_4,\vp)}} .
\]
As for the quark propagator in Landau gauge we expect also in Coulomb gauge
that the effective quark mass extracted from the static propagator is considerably smaller
than the one extracted from the four-dimensional propagator.

\section{Conclusions}
The gap equation of Ref.~\cite{Campagnari:2016wlt} has been rederived within the framework of the
canonical recursive Dyson--Schwinger equations. We have shown that the additional Dirac
structure in the bare quark-gluon vertex of the vacuum wave functional not only eliminates
the UV divergences from the gap equation (as shown already in Refs.~\cite{Vastag:2015qjd,Campagnari:2016wlt})
but is also crucial to ensure multiplicative renormalizability of the quark propagator.
We have performed a quenched semi-perturbative calculation assuming a bare quark-gluon
vertex. Unlike the covariant functional approaches in Landau gauge, where the dressing
of the (four-dimensional) quark-gluon vertex is crucial for obtaining spontaneous breaking
of chiral symmetry, in the present Hamiltonian approach the bare quark-gluon vertex in
the vacuum wave functional is sufficient to reproduce the phenomenological value of the
quark condensate. In the present approach the dominant IR contribution, which triggers
the spontaneous breaking of chiral symmetry, comes from the confining Coulomb potential.
We have also shown that, depending on the details of the momentum dependence, the
effective quark mass obtained in the Hamiltonian approach cannot be compared with the
(constituent) mass extracted from  the corresponding four-dimensional propagator and is
expected to be considerably smaller than the latter. The results obtained in the present
paper are quite encouraging for a fully self-consistent solution of the coupled
variational and CRDSEs.

\begin{acknowledgments}
The authors thank W.~Vogelsang for the LHAPDF evaluation of the coupling, E.~Ebadati for
providing the numerical solution of the gap equation, M.~Q.~Huber for the Landau gauge
quark propagator data, and M.~Quandt for a critical reading of the manuscript. This work
was supported but the Deutsche Forschungsgemeinschaft (DFG) under contract
No.~DFG-Re856/10-1.
\end{acknowledgments}


\appendix

\section{Coherent-state representation of fermion fields}
The coherent-state representation of the fermionic Fock space has been introduced in
Ref.~\cite{Campagnari:2015zsa} in coordinate space. For the sake of completeness we
collect here the relevant results in momentum space.
The Dirac field $\psi$ is expanded in the usual way
\begin{equation}\label{dir2a}
\begin{split}
\psi(\vx) &= \int \dfr[3]{p} \e^{\I*\vp\cdot\vx} \psi^m (\vp) , \\
\psi(\vp) &= \frac{1}{\tsqrt{2 E_\vp}} \bigl[ u (\vp,s) \, b(\vp,s) + v (-\vp,s) \, d^{\dag}(-\vp,s) \bigr]
\end{split}
\end{equation}
in terms of the eigenspinors $u(\vp,s)$, $v(\vp,s)$ of the free Dirac Hamiltonian $h_0(\vp)$ [\Eqref{dir}]
satisfying the eigenvalue equations
\[
h_0(\vp) \, u(\vp,s) = E_\vp \, u(\vp,s) , \qquad
h_0(\vp) \, v(-\vp,s) = - E_\vp \, v(-\vp,s) ,
\]
where $s = \pm 1$ accounts for the two spin degrees of freedom. With the usual 
normalization the Dirac eigenspinors satisfy the orthonormality relations
\begin{equation}\label{spinnorm}
\begin{gathered}
u^\dag(\vp,s) \, u(\vp,s') = 2 E_\vp \, \delta_{ss'} = v^\dag(\vp,s) \, v(\vp,s') , \\
u^\dag(\vp,s) \, \beta \, u(\vp,s') = 2 m\, \delta_{ss'} = -v^\dag(\vp,s) \, \beta \, v(\vp,s') ,  \\
u^\dag(\vp,s) \, v(-\vp,s') = 0 . 
\end{gathered}
\end{equation}
The expansion coefficients $b(\vp,s)$, $d^\dagger(\vp,s)$ are annihilation and creation operators 
satisfying the usual anti-com\-mu\-ta\-tion relations 
\[
\anticomm{b(\vp,s)}{b^{\dag}(\vq,t)} = \delta_{st} \, (2 \pi)^3 \delta(\vp-\vq) = \anticomm{d(\vp,s)}{d^{\dag}(\vq,t)} ,
\]
which, with the normalization (\ref{spinnorm}), ensure that the Fermi field in coordinate
space has the canonical anticommutation relation
\[
\{ \psi(\vx), \psi^{\dagger} (\vy) \} = \delta(\vx-\vy) .
\]
Furthermore, the operators $b(\vp,s)$ and $d(\vp,s)$ annihilate the filled Dirac sea of the
free fermions denoted by $\ket{0}$, i.e.
\[
b(\vp,s) \ket{0} = 0 = d(\vp,s) \ket{0} .
\]
The eigenspinors $u$ and $v$ are also eigenvalues of the projectors \Eqref{qptproj3}
\[
\begin{aligned}
\Lambda_{+}(\vp) \, u(\vp,s) &= u(\vp,s), \qquad & \Lambda_{+}(\vp) \, v(-\vp,s) &= 0 , \\
\Lambda_{-}(\vp) \, v(-\vp,s) &= v(-\vp,s), \qquad & \Lambda_{-}(\vp) \, u(\vp,s) &= 0 ,
\end{aligned}
\]
Furthermore, the projectors $\Lambda_{\pm}$ are related to the Dirac spinors by the
following completeness relations
\[
\sum_s \frac{u(\vp,s) \otimes u^\dag(\vp,s)}{2 E_\vp} = \Lambda_{+}(\vp), \qquad
\sum_s \frac{v(-\vp,s) \otimes v^\dag(-\vp,s)}{2 E_\vp} = \Lambda_{-}(\vp).
\]
Since we have two sets of fermion operators $b$, $b^\dag$ and $d$, $d^\dag$,
corresponding to particles and anti-particles, we need also two different sets of Grassmann variables.
Given the decomposition \Eqref{dir2a} of the Dirac field it is convenient
to define the coherent fermion states $\lvert\xi_+,\xi_-^*\rangle$ of the Dirac fermions by
\[
\begin{split}
b(\vp,s) \ket{\xi_+,\xi_-^*} &= \xi_+(\vp,s)\ket{\xi_+,\xi_-^*} , \\
d(\vp,s) \ket{\xi_+,\xi_-^*} &= \xi_-^{*}(\vp,s) \ket{\xi_+,\xi_-^*} ,
\end{split}
\]
and to introduce the Grassmann-valued Dirac spinor fields
\begin{equation}\begin{split}\label{spingrass}
\xi_+(\vp) &\coloneq \frac{1}{\tsqrt{2E_\vp}} \sum\limits_s \: u(\vp,s) \, \xi_+(\vp,s) , \\
\xi_-^{\dag}(\vp) &\coloneq \frac{1}{\tsqrt{2E_\vp}} \sum\limits_s \: v^\dag(-\vp,s) \, \xi_-^{*}(-\vp,s) ,
\end{split}
\end{equation}
which satisfy
\[
\Lambda_\pm(\vp) \, \xi_\pm(\vp) = \xi_\pm(\vp) .
\]
From Eqs.~\eqref{spinnorm} follow the inverse relations to \Eqref{spingrass}
\[
\begin{split}
\xi_+(\vp,s) &= \frac{1}{\tsqrt{2E_\vp}} \: u^\dag(\vp,s) \, \xi_+(\vp) , \\
\xi_-^{*}(\vp,s) &= \frac{1}{\tsqrt{2E_\vp}} \: \xi_-^{\dag}(\vp) v^\dag(-\vp,s) .
\end{split}
\]
For simplicity we will simply write $\ket{\xi}$ instead of $\ket{\xi_+,\xi_-^*}$.
With these definitions we find
\begin{equation}\label{dirrep}
\begin{aligned}
\frac{u^\dag(\vp,s)}{\tsqrt{2E_\vp}} \: \bra{\xi} b^{m\dag}(\vp,s) &= \xi_+^{m\dag}(\vp) \,\bra{\xi}\qquad&
\frac{u(\vp,s)}{\tsqrt{2E_\vp}} \: \bra{\xi} b^m(\vp,s) &= \frac{\delta}{\delta\xi_+^{m\dag}(\vp)} \, \bra{\xi} \\
\frac{v(-\vp,s)}{\tsqrt{2E_\vp}} \: \bra{\xi} d^{m\dag}(-\vp,s) &= \xi_-^m(\vp)\,\bra{\xi}  &
\frac{v^\dag(-\vp,s)}{\tsqrt{2E_\vp}} \: \bra{\xi} d^m(-\vp,s) &= \frac{\delta}{\delta\xi_-^m(\vp)}\, \bra{\xi}
\end{aligned}
\end{equation}
Furthermore, the coherent-state representation of a Fock state $\ket{\varPhi}$ of the Dirac
fermions is given by
\[
\varPhi[\xi_+^\dag,\xi_-] = \braket{\xi}{\varPhi}, \qquad
\varPhi^*[\xi_+,\xi_-^\dag] = \braket{\varPhi}{\xi} .
\]
In the following it will be also convenient to assemble the independent fields $\xi_+$ and
$\xi_-$ in a single Grassmann-valued spinor
\[
\xi(\vp) = \xi_+(\vp) + \xi_-(\vp), \qquad
\xi_\pm(\vp) = \Lambda_\pm (\vp) \, \xi(\vp) .
\]
In analogy to the Fourier decomposition (\ref{dir2a}) of the Fermi field we also introduce
the Grassmann fields in the coordinate representation
\[
\xi_\pm (\vx) = \int \dfr[3]{p} \ \e^{\I* \vp \cdot \vx} \, \xi_\pm (\vp) ,
\]
which implies
\[
\xi (\vx) = \xi_+ (\vx) + \xi_- (\vx) .
\]
and
\[
\frac{\delta}{\delta \xi_\pm (\vx)} = \int \dfr[3]{p} \: \e^{-\I* \vp \cdot \vx} \frac{\delta}{\delta \xi_\pm (\vp)} \, .
\]
From Eqs.~(\ref{dirrep}) then follows that the action of the Fermi-field $\psi(\vx)$ [Eq.~(\ref{dir2a})] on the 
coherent state $\ket{\xi} \equiv \ket{\xi_+ , \xi^*_-}$ is given by
\[
\begin{split}
\bra{\xi} \psi (\vx) & = \left( \xi_-(\vx) + \frac{\delta}{\delta \xi^\dagger_+(\vx)} \right) \bra{\xi} , \\
\bra{\xi} \psi^\dagger (\vx) &= \left( \xi^\dagger_+ (\vx) + \frac{\delta}{\delta \xi_- (\vx)} \right) \bra{\xi} .
\end{split}
\]


\section{The Quark CRDSE}
\label{app:qdse}

In the bare vertex approximation the CRDSE~\eqref{1010-17a} for the quark propagator
\Eqref{qdse5a} reduces in the chiral limit to the following set of equations for the dressing functions
\begin{subequations}\label{qdse6}
\begin{align}
\label{qdse6a}
A_p &= 1 + 
\begin{multlined}[t]
\frac{g^2 C_F}{2} \int \dfr[3]{q} \: \frac{1}{\Omega(\vp+\vq) \, \Delta_q}
\biggl\{ 
 A_q \biggl[ X_-(\vp,\vq) \, \frac{\abs{V(\vp,\vq)}^2 + \abs{V(\vq,\vp)}^2}{2}
  + X_+(\vp,\vq) \, \frac{\abs{W(\vp,\vq)}^2 + \abs{W(\vq,\vp)}^2}{2} \biggr] \\
 + D_q \biggl[ X_-(\vp,\vq) \, \frac{\abs{V(\vp,\vq)}^2 - \abs{V(\vq,\vp)}^2}{2}
  + X_+(\vp,\vq) \, \frac{\abs{W(\vp,\vq)}^2 - \abs{W(\vq,\vp)}^2}{2} \biggr]
\biggr\} 
\end{multlined} 
\displaybreak[0] \\
\label{qdse6b}
B_p &= \Re s_p +
\frac{g^2 C_F}{2} \int \dfr[3]{q} \: \frac{1}{\Omega(\vp+\vq) \, \Delta_q}
\begin{multlined}[t]
\Bigl\{  B_q \Bigl[ X_-(\vp,\vq) \Re\bigl[ V(\vp,\vq) \, V(\vq,\vp) \bigr] - X_+(\vp,\vq) \Re\bigl[ W(\vp,\vq) \, W(\vq,\vp) \bigr] \Bigr] \\
 + C_q \Bigl[ X_-(\vp,\vq) \Im\bigl[ V(\vp,\vq) \, V(\vq,\vp) \bigr] - X_+(\vp,\vq) \Im\bigl[ W(\vp,\vq) \, W(\vq,\vp) \bigr] \Bigr]
\Bigr\} ,
\end{multlined}
\displaybreak[0] \\
\label{qdse6c}
C_p &= \Im s_p +
\frac{g^2 C_F}{2} \int \dfr[3]{q} \: \frac{1}{\Omega(\vp+\vq) \, \Delta_q}
\begin{multlined}[t]
\Bigl\{ - C_q \Bigl[ X_-(\vp,\vq) \Re\bigl[ V(\vp,\vq) \, V(\vq,\vp) \bigr] - X_+(\vp,\vq) \Re\bigl[ W(\vp,\vq) \, W(\vq,\vp) \bigr] \Bigr] \\
 + B_q \Bigl[ X_-(\vp,\vq) \Im\bigl[ V(\vp,\vq) \, V(\vq,\vp) \bigr] - X_+(\vp,\vq) \Im\bigl[ W(\vp,\vq) \, W(\vq,\vp) \bigr] \Bigr]
\Bigr\} ,
\end{multlined}
\displaybreak[0] \\
\label{qdse6d}
D_p &= \frac{g^2 C_F}{2} \int \dfr[3]{q} \: \frac{1}{\Omega(\vp+\vq) \, \Delta_q}
\begin{multlined}[t]
\biggl\{ 
 D_q \biggl[ X_-(\vp,\vq) \, \frac{\abs{V(\vp,\vq)}^2 + \abs{V(\vq,\vp)}^2}{2}
  + X_+(\vp,\vq) \, \frac{\abs{W(\vp,\vq)}^2 + \abs{W(\vq,\vp)}^2}{2} \biggr] \\
 +A_q \biggl[ X_-(\vp,\vq) \, \frac{\abs{V(\vp,\vq)}^2 - \abs{V(\vq,\vp)}^2}{2}
  + X_+(\vp,\vq) \, \frac{\abs{W(\vp,\vq)}^2 - \abs{W(\vq,\vp)}^2}{2} \biggr]
\biggr\} .
\end{multlined}
\end{align}
\end{subequations}
where $\Delta_q$ is given by \Eqref{qdse8} and $X_\pm(\vp,\vq)$ by \Eqref{qdse7}.


\section{\label{subsec:vebv}The vacuum energy density}

When the full quark-gluon vertex $\bar\Gamma$ [\Eqref{qgv}] is replaced by the bare one $\bar\Gamma_0$
[\Eqref{bqgv}, \eqref{qgv0}], the remaining traces in the energy density contributions
Eqs.~\eqref{eD2} and \eqref{eE7} can be worked out explicitly. One finds for the second
piece of the single-particle Hamiltonian \Eqref{eD2}
\begin{multline}\label{lve1bis}
e^{(1)}_\mathrm{D} = 
g^2 C_F \int \dfr[3]{q} \dfr[3]{\ell} \,
\frac{X_-(\vq,\vl)}{\Delta_q \, \Delta_\ell \, \Omega(\vq+\vl)}
  \begin{aligned}[t]
  \Bigl\{
  &\bigl[ \bigl( A_q - D_q \bigr) \bigl( A_\ell + D_\ell \bigr) + B_q\,B_\ell - C_q\,C_\ell \bigr] \Re V(\vq,\vl)\\
  &+ \bigl[ B_q\,C_\ell + C_q\,B_\ell \bigr] \Im V(\vq,\vl) \Bigr\}
  \end{aligned}
\\
\begin{aligned}[b]
{} + g^2 C_F \int \dfr[3]{q} \dfr[3]{\ell} \,
\frac{X_+(\vq,\vl)}{\Delta_q \, \Delta_\ell \, \Omega(\vq+\vl)} 
  \Bigl\{
  &\bigl[ \bigl( A_q - D_q \bigr) B_\ell + \bigl( A_\ell + D_\ell \bigr) B_q\bigr] \Re W(\vq,\vl) \\
  &+\bigl[ \bigl( A_q - D_q \bigr) C_\ell + \bigl( A_\ell + D_\ell \bigr) C_q\bigr] \Im W(\vq,\vl) \Bigr\}
  \end{aligned}
\end{multline}
with $X_\pm(\vq,\vl)$ given in \Eqref{qdse7}. Furthermore, the contribution \Eqref{eE7} from the kinetic energy
of the gluons reduces to
\[
e^{}_E = g^2 C_F \int \dfr[3]{q} \dfr[3]{\ell}
\, \frac{A_q+D_q}{\Delta_q}\frac{A_\ell+D_\ell}{\Delta_\ell} \Bigl\{ X_-(\vq,\vl) \abs{V(\vq,\vl)}^2 + X_+(\vq,\vl) \abs{W(\vq,\vl)}^2 \Bigr\} .
\]


\section{\label{app:gap}The Quark Gap Equation}

In the bare vertex approximation one finds from the minimization of the energy density
for the scalar kernel $s_p$ the following equation:
\begin{equation}\label{x1}
\begin{aligned}[b]
\abs\vp s_p
={}& \frac{g^2 C_F}{2}\int\dfr[3]{q} \frac{1}{(1+s_q^2) \Omega(\vp+\vq)} \\
&\times\biggl\{\!
\begin{aligned}[t]
&X_-(\vp,\vq) \, V(\vp,\vq) \bigl[ (1-s_p^2)s_q - 2s_p) \bigr] + X_+(\vp,\vq) \, W(\vp,\vq) \bigl[1-s_p^2-2s_ps_q \bigr] \\
& - \frac{\abs\vp}{1+s_p^2}
 \begin{aligned}[t]
 &\Bigl[ X_-(\vp,\vq) \, V^2(\vp,\vq) \bigl[ s_p(s_p^2-3)+s_q(1-3s_p^2) \bigr] \\
 &\qquad+ X_+(\vp,\vq) \, W^2(\vp,\vq)\bigl[ s_p(s_p^2-3) - s_q(1-3s_p^2) \bigr] \Bigr]
 \end{aligned}
\\
&- \frac{\abs{\vq}}{1+s_q^2}
 \begin{aligned}[t]
 & \Bigl[ X_-(\vp,\vq) \, V^2(\vp,\vq) \bigl[ (1-s_p^2)s_q - s_p(1-s_q^2) \bigr] \\
 &\qquad- X_+(\vp,\vq) \, W^2(\vp,\vq) \bigl[(1-s_p^2)s_q+s_p(1-s_q^2) \bigr] \Bigr] \biggr\}
 \end{aligned}
\end{aligned} \\
&+ \frac{g^2 C_F}{2} \int\dfr[3]{q} \, \frac{s_p}{1+s_q^2} \bigl[ X_-(\vp,\vq) \, V^2(\vp,\vq) + X_+(\vp,\vq) \, W^2(\vp,\vq) \bigr] \\
&+ \frac{g^2 C_F}{2} \int\dfr[3]{q} \frac{F(\vp-\vq)}{1+s_q^2} \bigl[ s_q (1-s_p^2) - \uvp\cdot\uvq (1-s_p^2) \bigr]
\end{aligned}
\end{equation}


\bibliographystyle{h-physrev5}
\bibliography{biblio-spires}

\end{document}